\documentclass[usenatbib]{mn2e}
\usepackage[pdftex]{color}
\usepackage[draft]{graphicx}
\usepackage[draft]{hyperref}
\usepackage[fleqn]{amsmath}     
\usepackage{amssymb}

\usepackage{longtable}
\usepackage{lscape}

\title[Morphology and kinematics]{Morphology and kinematics of orbital components in CALIFA galaxies across the Hubble sequence}
\author[Zhu et al]{Ling Zhu$^1$\thanks{E-mail: lzhu@mpia.de} , Glenn van de Ven$^{2,1}$, Jairo M\'endez-Abreu$^{3,4}$, Aura Obreja$^5$.
\\
  $^1$ Max Planck Institute for Astronomy, K\"onigstuhl 17, 69117 Heidelberg, Germany \\
  $^2$ European Southern Observatory, Karl-Schwarzschild-Str. 2, 85748 Garching b. M\"unchen, Germany\\
  $^3$ Instituto de Astrofisica de Canarias C/ Via Lactea, s/n, E-38205, La Laguna, Tenerife, Spain\\
  $^4$ Departamento de Astrof\'isica, Universidad de La Laguna, E-38200 La Laguna, Tenerife, Spain\\
  $^5$ University Observatory Munich, Scheinerstr. 1, D-81679 Munich, Germany}

\begin{document}
\date{}
\maketitle

\begin{abstract}
Based on the stellar orbit distribution derived from orbit-superposition Schwarzschild models, we decompose each of 250 representative present-day galaxies into four orbital components: cold with strong rotation, warm with weak rotation, hot with dominant random motion and counter-rotating (CR).
We rebuild the surface brightness ($\Sigma$) of each orbital component and we present in figures and tables a quantification of their morphologies using the Sersic index \textit{n}, concentration $C = \log{(\Sigma_{0.1R_e}/\Sigma_{R_e})}$ and intrinsic flattening $q_{\mathrm{Re}}$ and $q_{\mathrm{Rmax}}$, with $R_e$ the half-light-radius and $R_{\mathrm{max}}$ the CALIFA data coverage. 
We find that: (1) kinematic hotter components are generally more concentrated and rounder than colder components, and (2) all components become more concentrated and thicker/rounder in more massive galaxies; they change from disk-like in low mass late-type galaxies to bulge-like in high-mass early type galaxies. Our findings suggest that Sersic \textit{n} is not a good discriminator between rotating bulges and non-rotating bulges.
The luminosity fraction of cold orbits $f_{\rm cold}$ is well correlated with the photometrically-decomposed disk fraction $f_{\rm disk}$ as $f_{\mathrm{cold}} = 0.14 + 0.23f_{\mathrm{\mathrm{disk}}}$. Similarly, the hot orbit fraction $f_{\rm hot}$ is correlated with the bulge fraction $f_{\rm bulge}$ as $f_{\mathrm{hot}} = 0.19 + 0.31f_{\mathrm{\mathrm{bulge}}}$. The warm orbits mainly contribute to disks in low-mass late-type galaxies, and to bulges in high-mass early-type galaxies.  
The cold, warm, and hot components generally follow the same morphology ($\epsilon = 1-q_{\rm Rmax}$) versus kinematics ($\sigma_z^2/\overline{V_{\mathrm{tot}}^2}$) relation as the thin disk, thick disk/pseudo bulge, and classical bulge identified from cosmological simulations. 
\end{abstract}


\section{Introduction}
\label{S:intro}
The distribution of stellar orbits in a galaxy determines the galaxy's morphology and kinematic properties. For example, stars on high-circularity orbits form fast-rotating, flat disks, while stars on low-circularity orbits form slow-rotating, round bulges. 
To infer the properties of different galaxy components is, however, far from trivial, because their photometry and kinematics
are blended together. 

Historically, a bulge is defined as a bright central concentration \citep{Hubble1936}, and this definition has been extensively used for a visual classification of galaxies (e.g. \citealt{Buta2013}). Galaxies have increasing bulges from spirals to ellipticals on the Hubble sequence.  
IFU surveys: like SAURON \citep{Davies2001}, ATLAS3d \citep{Cappellari2011}, CALIFA \citep{CALIFA12}, SAMI \citep{Croom2012}, MaNGA \citep{Bundy2015}, provide stellar kinematic maps for a large number of galaxies. The mean velocity and velocity dispersion maps can be used to quantify the overall projected ordered-to-random motion ratio, through the spin parameter $\lambda_{\rm R_e}$ \citep{Emsellem2011}. Fast-rotating systems with low velocity dispersion have $\lambda_{\rm R_e} \sim 1$, while slow rotating systems dominated by random motions have $\lambda_{\rm R_e} \sim 0$.
In the diagram of $\lambda_{\rm R_e}$ vs. ellipticity $\epsilon$, late-type galaxies with smaller bulges generally lie in the area with high $\lambda_{\rm R_e}$ and $\epsilon$, while early-type galaxies with larger bulges lie in the area with low $\lambda_{\rm R_e}$ and $\epsilon$ \citep{Emsellem2011, Cappellari2016}.
The correlation is strong with $\lambda_{\rm R_e}$ and galaxy's intrinsic flattening \citep{Foster2017}.

However, to quantify the structure and dynamics of the galaxy components and so constrain their assembly history, 
one needs to decompose the galaxies in a physical way. 

A photometric bulge-disk decomposition is a straight-forward way to start. The images of galaxies are fitted by a combination of a central Sersic bulge \citep{Sersic1968} and an exponential outer disk \citep{Freeman1970}. 
In the observed stellar kinematic maps, bulge- and disk-dominated regions show signs of lower (higher) and higher (lower) mean velocity (velocity dispersion), respectively.
However, the contributions of bulge and disk to the observed line-of-sight kinematic maps are not linear, and it is very challenging to obtain the kinematic properties of bulge and disk separately \citep{Tabor2017}. 
Even in the central region, where the bulge dominates the galaxy light, it is hard to correct for contribution by an underlying disk \citep{Mendez-Abreu2018}.
Moreover, photometric decomposition is model-dependent and it can be subject to large uncertainties.

On the other hand, decomposition based on the stellar orbits' circularity distribution has been successfully applied to simulated galaxies and shown to be powerful in identifying different components with well-defined morphology and kinematics \citep{Abadi2003, Scannapieco2010, Martig2012, Obreja2016, Obreja2018a}.  
This results in a complicated relation between morphology and kinematics of components: the high angular momentum orbits do not necessarily form an exponential disk, and bulges show a large variety of flattening and rotation.

Recently we present in \citet{Zhu2018b} the orbit distribution of 300 CALIFA galaxies obtained via Schwarzschild orbit-superposition models. In this paper, we use these orbital distributions to decompose these galaxies into four components and infer the photometry and kinematics of each component. In Section~\ref{S:sample}, we describe the sample and show the decomposition.
In Section~\ref{S:morphology}, we quantify the intrinsic morphological properties of the four orbital components. In Section~\ref{S:observation}, we compare these morphological properties with the results from photometric studies. In Section~\ref{S:k-m}, we investigate the morphology versus kinematics relation. We discuss the results in Section~\ref{S:discussion} and conclude in Section~\ref{S:summary}.

\section{Orbital components}
\label{S:sample}
\subsection{Galaxy sample and Schwarzschild models}
\label{SS:sample}
\citet{califakin2017} presents stellar kinematic maps of 300 galaxies from the CALIFA survey (see \citealt{CALIFA12}).
These 300 galaxies encompasses the main morphological galaxy types with a total stellar mass, $M_*$, between $10^{8.7}$ to $10^{11.9} \, M_{\odot}$, and a well-defined selection function between $10^{9.7}$ to $10^{11.4} \, M_{\odot}$ \citep{Walcher2014}. Of the 300 galaxies, 279 are identified as isolated and we further exclude 19 objects that could be biased by dust lanes. The remaining 260 galaxies are suitable for dynamical modelling. 
We focus on the 250 galaxies with kinematic data coverage $R_{\mathrm{max}} > R_{\rm e}$, with $R_e$ the half-light radius.

We created orbit-superposition Schwarzschild models \citep{vdB2008} which simultaneously fit the observed surface brightness and stellar kinematics for each galaxy \citep{Zhu2018a}. 
Orbits are characterized by two main properties: the time-averaged radius $r$ representing the size of each orbit, and circularity, $\lambda_z\equiv J_z/J_{\mathrm{max}}(E)$ around the short $z$ axis, normalised by the maximum of a circular orbit with the same binding energy $E$. Circular orbits have $\lambda_z=1$, radial or box orbits have $\lambda_z=0$, and counter-rotating orbits have $\lambda_z<0$. 
The stellar orbit distribution of the best-fit Schwarzschild model is then described by the probability density of orbit weights, $p(\lambda_z, r)$.  The overall orbit circularity distribution, $p(\lambda_z)$, is obtained by integrating $p(\lambda_z, r)$ over all $r < R_{\rm e}$.

We derived in \citet{Zhu2018b} a statistically representative stellar orbit distribution of present-day galaxies based on these 250 galaxies. Though about half of the 250 galaxies are barred, we tested also that the overall orbit's circularity distribution is not biased.

\subsection{Orbital decomposition}
\label{SS:decomposition}
The technique of orbital decomposition has been described in detail in \citet{Zhu2018a}. 
Given the overall circularity distribution, $p(\lambda_z)$, we divide the orbits within $1\,R_e$ into four components: cold ($0.8\leq \lambda_z\leq 1$), warm ($0.25<\lambda_z<0.8$), hot ($-0.25 \leq \lambda_z \leq 0.25$) and counter-rotating (CR; $\lambda_z < -0.25$). 
We denote the galaxy's stellar orbit distribution within $1\,R_e$ as luminosity fractions of four components: $f_{\mathrm{cold}}$, $f_{\mathrm{warm}}$, $f_{\mathrm{hot}}$, and $f_{\mathrm{CR}}$. The separation and orbit fractions are the same as in \citet{Zhu2018b}. The uncertainties of the luminosity fractions are obtained by testing against 131 mock data sets created from a variety of simulations. The systematic bias, systematic errors and statistical errors for low-mass spiral galaxies, high-mass spiral galaxies, lenticulars and ellipticals are listed in \citet{Zhu2018b}. Here we assign errors to $f_{\mathrm{cold}}$, $f_{\mathrm{warm}}$, $f_{\mathrm{hot}}$, and $f_{\mathrm{CR}}$ of each galaxy according to the galaxy type.

\begin{figure}
\centering\includegraphics[width=8.0cm]{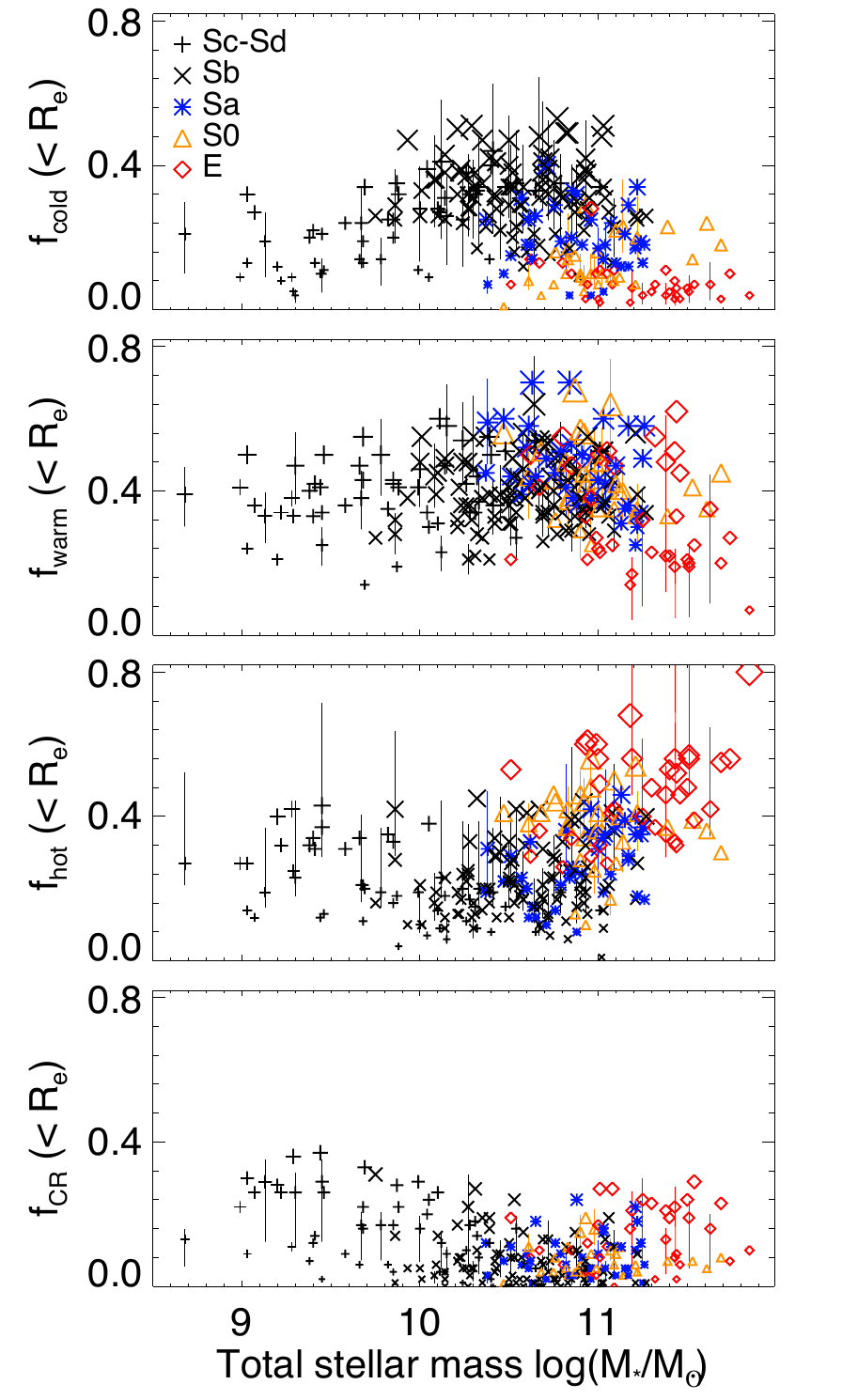}
\caption{Orbit fractions as function of galaxy's total stellar mass $M_*$. From top to bottom; cold, warm, hot and CR orbital fractions ($f_{\mathrm{cold}}, f_{\mathrm{warm}}, f_{\mathrm{hot}}, f_{\mathrm{CR}}$). Black pluses, black crosses, blue asterisks, orange triangles and red diamonds represent Sc-Sd, Sb, Sa, S0 and E galaxies, with symbol sizes indicating the luminosity fractions. The short vertical lines indicate the $1\,\sigma$ uncertainties, including both statistical uncertainties as well as systematic biases and uncertainties as inferred from tests with simulated galaxies \citep{Zhu2018b}, we have errors for all, but only shown that for randomly $1/5$ of the points. The change of orbital distributions maps to change of galaxy's Hubble types. The data points in this figure are given in Table~\ref{tab:orbitf} in the appendix. }
\label{fig:M_f}
\end{figure}

In Figure~\ref{fig:M_f}, we show the luminosity fraction of these four components as a function of stellar mass, with pluses, crosses, asterisks, triangles and diamonds representing Sc/Sd, Sb, Sa, S0 and elliptical galaxies, respectively.
We use the total stellar mass $M_*$ derived by fitting the spectral energy distributions (SEDs) from multi-band photometry using a linear combination of single stellar population synthetic spectra of different ages and metallicities, adopting a \citet{Kroupa2001} IMF \citep{Walcher2014}.

For galaxies with $M_{*} > 10^{10}\, M_{\odot}$, the majority of galaxies change from late type to early type (Sb to Sa, to S0, to ellipticals) with decreasing cold component and increasing hot component.  
However, the Hubble type does not reflect all about the orbit distribution, galaxies classified as the same Hubble type still have significant variation of stellar orbits, for instance, the cold component still decrease  (hot component increases) with increasing $M_{*}$ among the galaxies classified as Sa, similar to S0 and ellipticals. The warm component keeps almost constant and only drops in the most massive elliptical galaxies. CR orbit fraction is small in most galaxies, and it becomes larger in the least-massive Sc/Sd galaxies and the most-massive elliptical galaxies.


Each of the four components spans a certain range of circularity $\lambda_z$, here we calculate the average circularity of each component.
The average $\lambda_z$ of cold, warm, hot and CR components are $0.92\pm0.02$, $0.54\pm0.05$, $0.02\pm0.04$, $-0.53\pm0.12$, respectively. 
The errors represent the $1\sigma$ scatter among the whole sample of 250 galaxies.
The $\lambda_z$ of each component is quite uniform; there is not notable systematic change with total stellar mass $M_*$, or with Hubble type.

\subsection{Reconstruction of orbital components}
\label{SS:4comps}

\begin{figure*}
\centering\includegraphics[width=9.0cm]{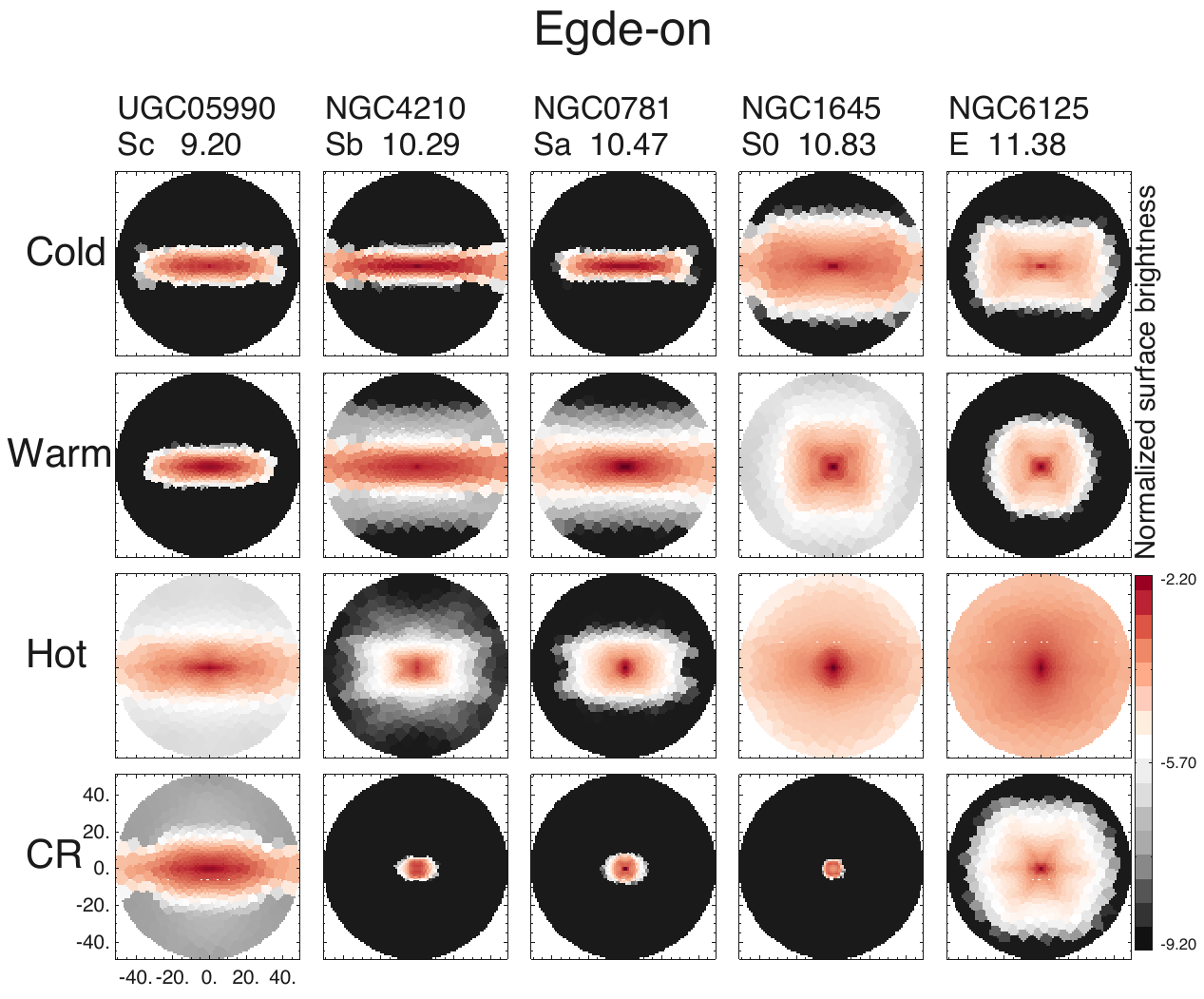}\centering\includegraphics[width=8.7cm]{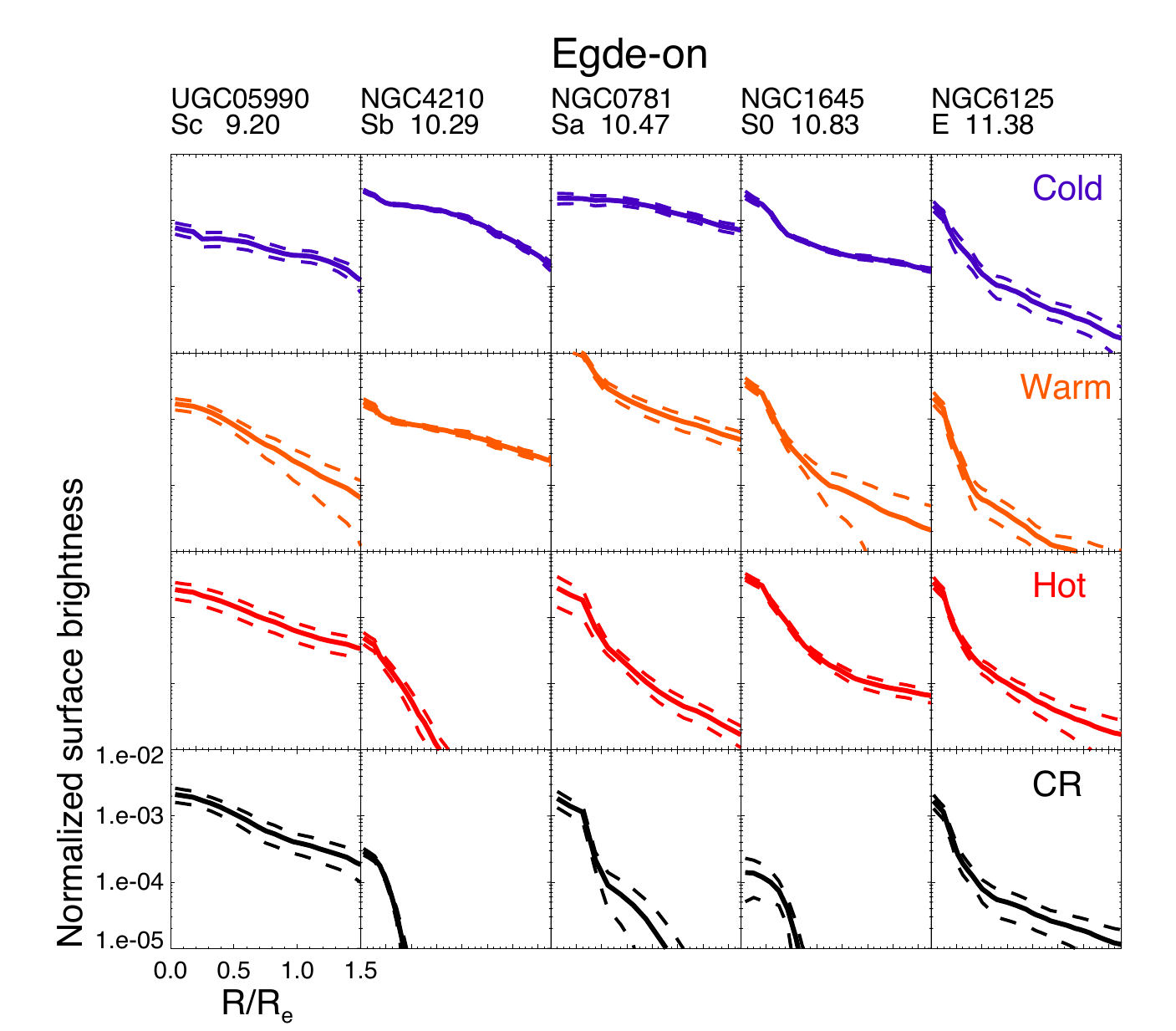}
\caption{The edge-on projected surface brightness of the cold, warm, hot, and CR components for five representative galaxies. The name of the galaxies are listed at the top of columns, as well as their Hubble types and their total stellar masses as $\log(M_*/M_{\odot})$. In the left panels, we show the 2D surface brightness of the best-fitting model. The colorbars indicate the values of the normalized surface brightness in $\log(L/{\rm arcsec}^2)$, with the same scale for all panels, the total luminosity within $1\,R_e$ of each galaxy is normalized to unity. All panels share the same $x$ and $y$ axis range from $-50$ to $50$ arcsec, as shown in the bottom-left panels. In the right panels, we show the mean (solid line) and $1\,\sigma$ scatter (dashed lines) of the 1D SB along major axis from the models within $1\,\sigma$ confidence level, with x axis in $R/R_e$, y axis in the same unit as the density maps on the left. }
\label{fig:maps5}
\end{figure*}

An advantage of the orbit-superposition model is that, using the orbits, we can reconstruct the 3D density distribution and kinematic properties of any component based on the orbital decomposition \citep{Zhu2018a}. We reconstruct the density distribution for the cold, warm, hot, and CR components of all 250 galaxies. 

We illustrate the projected edge-on surface brightness (SB) with five galaxies from low-mass late-type to high-mass early-type in Figure~\ref{fig:maps5}. At the top we list the galaxy names, their Hubble types, and total stellar mass as $\log(M_{*}/M_{\odot})$. 
2D maps of the best-fitting model of each galaxy are shown in the left panels. The four panels from top to bottom are the surface brightness of the cold, warm, hot, and CR components. The total luminosity within $1\,R_e$ of each galaxy is normalized to unity. We also select the models within the $1\sigma$ confidence level \citep{Zhu2018a}, there are 350, 25, 49, 36, and 120 models within $1\sigma$ for UGC05990, NGC4210, NGC0781, NGC1645 and NGC6125, respectively. The mean and $1\sigma$ scattering of the four components' surface brightness are shown as the solid and dashed curves in the right panels. There is only mild variation of the morphology of each orbital component within the $1\sigma$ confidence level.

These five galaxies represent typical cases with the variaty of their four components' morphology showcasing the systematic changes we will describe in following sections.
The edge-on projected maps reveal the intrinsic flattening. UGC05990 is a Sc galaxy with $M_* = 10^{9.2}\, M_{\odot}$; its four components are intrinsically flat and radially extended. NGC 4210 which is a Sb galaxy with $M_* = 10^{10.29}\, M_{\odot}$ and NGC0781 which is a Sa galaxy with $M_* = 10^{10.47}\, M_{\odot}$ are similar; the four components become significantly thicker and more concentrated from cold, warm to hot component, while the CR components are only small fractions and more concentrated. 
In the S0 galaxy NGC1645 with $M_* = 10^{10.83}\, M_{\odot}$, all components become rounder compared to the Sb and Sa galaxies. This is also the case for the giant elliptical galaxy NGC6125 with $M_* = 10^{11.38}\, M_{\odot}$, which has a significant CR component with a round morphology.


\section{Morphology}
\label{S:morphology}

\subsection{Radial surface brightness profiles}
\label{SS:SB}

\begin{figure*}
\centering\includegraphics[width=16cm]{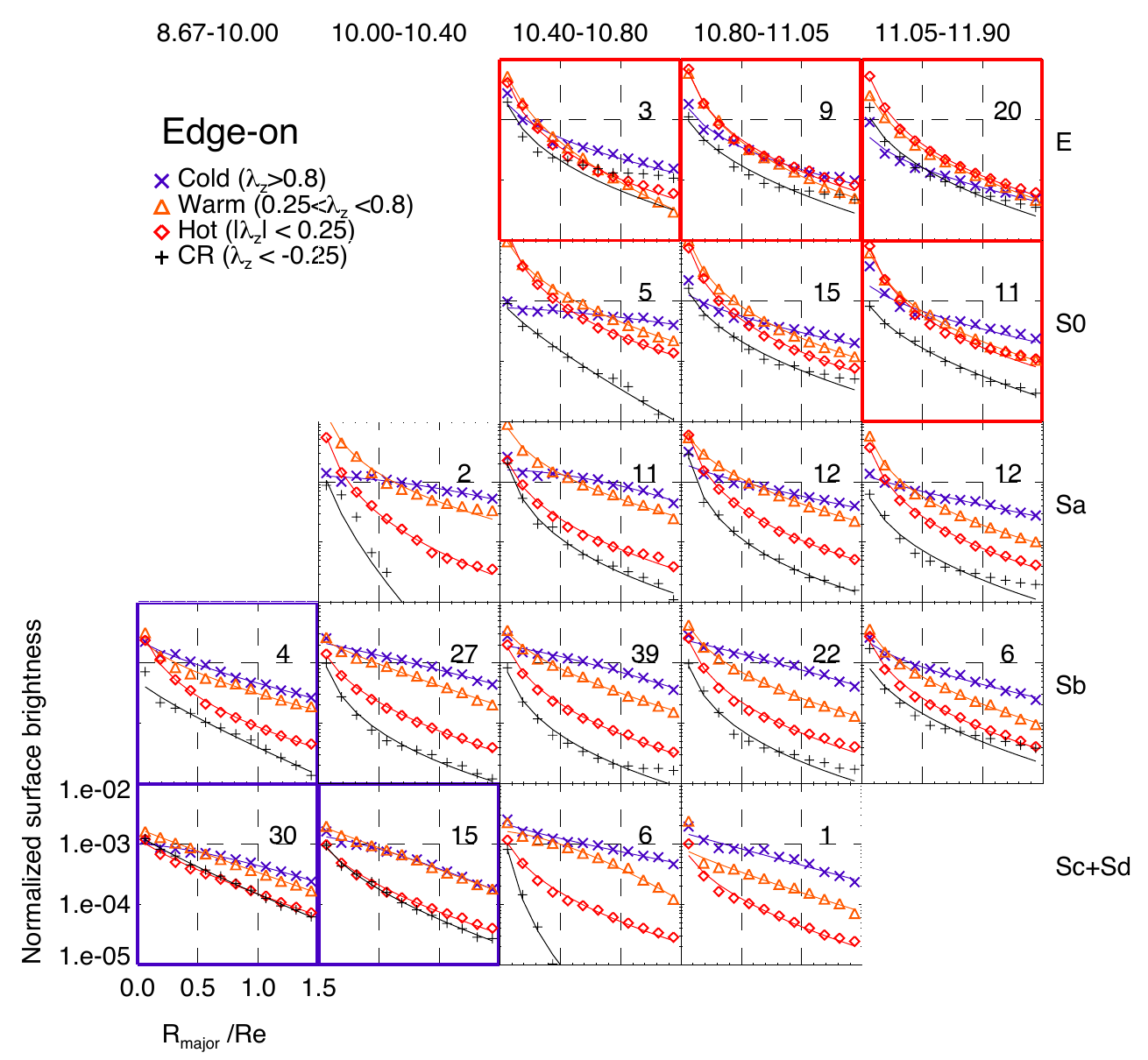}
\caption{Normalized edge-on surface brightness profile along the galaxy's major axis.
The 250 galaxies are divided into several bins according to their Hubble types and total stellar masses as $\log(M_*/M_{\odot})$, with the number of galaxies indicated in each bin. They have increasing stellar masses from left to right, and change from late-type to early type galaxies from bottom to top. The five galaxies shown in Figure~\ref{fig:maps5} lie in the five diagonal panels. In each panel, the blue crosses, orange triangles, red diamonds, and black pluses are the averaged SB of the cold, warm, hot, and CR components of these galaxies in that bin. The thin curves with corresponding colors are fitted Sersic profiles with parameters given in Table~\ref{tab:sb300}.}
\label{fig:sb300}
\end{figure*}

To have an overview of the whole sample, we divide the 250 galaxies into bins according to their total stellar mass and Hubble types. The five stellar mass bins are set with boundaries in $\log(M_*/M_{\odot})$ of $8.67, 10.0, 10.4, 10.8, 11.05, 11.9$ and the five Hubble classes are Sc+Sd, Sb, Sa, S0 and E. This results in the 19 panels in Figure~\ref{fig:sb300} with the number of galaxies indicated.

We show the four components' average edge-on surface brightness radial profiles. The five galaxies shown in Figure~\ref{fig:maps5} belong to the five diagonal panels from bottom-left to top-right. 
In each panel, the blue crosses, orange triangles, red diamonds and black pluses are the surface brightness profiles of the cold, warm, hot and CR components. In the middle panels, the cold components are radially extended, the warm components become more concentrated, the hot and CR components are the most concentrated ones. CR components only contribute small fractions and usually decline fast with radius. However, in the low-mass late-type galaxies, the three panels in the bottom-left (in blue), the four components have similar radial profiles and are all extended. In the high-mass early-type galaxies, the four panels in the top-right (in red), the four components also have similar radial profiles but are all concentrated.

\begin{table*}
\caption{The luminosity fractions and Sersic-fit parameters of cold, warm, hot, and CR components, for galaxies divided into 19 panels with 5 bins in stellar mass and 5 bins in Hubble types, as shown in Figure~\ref{fig:sb300}. 
The first two columns are Hubble type and stellar mass range of each bin, the four rows under each bin are five parameters of cold, warm, hot, and CR components from top to bottom, these five parameters are luminosity fraction within $R_e$ $f(<R_e)$, luminosity fraction within 5 kpc $f(<5$kpc), effective surface density $\Sigma_{\rm e}$ in unit of $10^3 L/{\rm arcsec}^2$, with the total luminosity of the galaxy within $R_e$ normalised to unity, effect radius $r_s$ in units of $R_e$ of the galaxy, and Sersic index \textit{n}. The corresponding Sersic fitting parameters for each single galaxy are included in the online material.}
\label{tab:sb300}\tiny 
\begin{tabular}{*{7}{l}}
\hline
\hline
Type & $\log(M_{*}/M_{\odot})$  & $f(<R_e)$ & $f(<5$kpc) & $\Sigma_e$ &$r_s/R_e$ &$n$  \\
\hline
     E   &     10.40-10.80   & 0.11 & 0.17 &0.117 & 1.50 & 1.74 \\
         &                   & 0.37 & 0.31 &0.457 & 0.48 & 2.13 \\
         &                   & 0.39 & 0.42 &0.103 & 1.00 & 3.72 \\
         &                   & 0.12 & 0.11 &0.133 & 1.00 & 0.73 \\
     E   &     10.80-11.05   & 0.10 & 0.09 &0.075 & 1.50 & 2.03 \\
         &                   & 0.35 & 0.37 &0.229 & 0.71 & 3.31 \\
         &                   & 0.45 & 0.44 &0.138 & 1.00 & 3.90 \\
         &                   & 0.10 & 0.10 &0.059 & 1.00 & 2.03 \\
     E   &     11.05-11.90   & 0.06 & 0.05 &0.039 & 1.50 & 1.56 \\
         &                   & 0.31 & 0.30 &0.118 & 0.93 & 2.46 \\
         &                   & 0.48 & 0.49 &0.116 & 1.00 & 3.82 \\
         &                   & 0.15 & 0.16 &0.053 & 1.00 & 2.35 \\
         \hline
    S0   &     10.40-10.80   & 0.07 & 0.15 &0.374 & 1.50 & 0.50 \\
         &                   & 0.45 & 0.43 &0.455 & 0.99 & 2.36 \\
         &                   & 0.42 & 0.38 &0.259 & 1.00 & 4.46 \\
         &                   & 0.06 & 0.05 &0.124 & 0.56 & 1.24 \\
    S0   &     10.80-11.05   & 0.12 & 0.14 &0.166 & 1.50 & 1.18 \\
         &                   & 0.44 & 0.43 &0.228 & 1.02 & 3.78 \\
         &                   & 0.36 & 0.35 &0.141 & 1.00 & 4.62 \\
         &                   & 0.08 & 0.09 &0.069 & 1.00 & 2.30 \\
    S0   &     11.05-11.90   & 0.17 & 0.16 &0.186 & 1.50 & 1.30 \\
         &                   & 0.41 & 0.39 &0.185 & 1.07 & 3.06 \\
         &                   & 0.35 & 0.37 &0.167 & 1.00 & 3.86 \\
         &                   & 0.07 & 0.07 &0.056 & 1.00 & 1.91 \\
         \hline
    Sa   &     10.00-10.40   & 0.16 & 0.22 &0.515 & 1.45 & 0.61 \\
         &                   & 0.52 & 0.53 &0.222 & 1.50 & 3.69 \\
         &                   & 0.25 & 0.19 &0.254 & 0.54 & 3.60 \\
         &                   & 0.07 & 0.07 &0.237 & 0.21 & 1.53 \\
    Sa   &     10.40-10.80   & 0.20 & 0.19 &0.719 & 1.15 & 0.56 \\
         &                   & 0.50 & 0.51 &0.255 & 1.50 & 2.58 \\
         &                   & 0.20 & 0.21 &0.076 & 1.00 & 2.90 \\
         &                   & 0.09 & 0.09 &0.053 & 0.76 & 4.06 \\
    Sa   &     10.80-11.05   & 0.17 & 0.19 &0.337 & 1.50 & 0.99 \\
         &                   & 0.46 & 0.46 &0.250 & 1.37 & 2.22 \\
         &                   & 0.28 & 0.27 &0.107 & 1.00 & 4.56 \\
         &                   & 0.08 & 0.09 &0.032 & 1.00 & 6.50 \\
    Sa   &     11.05-11.90   & 0.19 & 0.16 &0.240 & 1.50 & 0.96 \\
         &                   & 0.40 & 0.42 &0.091 & 1.50 & 3.71 \\
         &                   & 0.32 & 0.32 &0.078 & 1.00 & 3.96 \\
         &                   & 0.09 & 0.11 &0.024 & 1.00 & 2.50 \\
         \hline
    Sb   &      8.67-10.00   & 0.32 & 0.32 &0.261 & 1.45 & 1.30 \\
         &                   & 0.31 & 0.31 &0.161 & 1.50 & 1.68 \\
         &                   & 0.24 & 0.25 &0.086 & 1.00 & 2.91 \\
         &                   & 0.13 & 0.12 &0.067 & 0.76 & 1.18 \\
    Sb   &     10.00-10.40   & 0.34 & 0.34 &0.480 & 1.40 & 0.93 \\
         &                   & 0.39 & 0.40 &0.345 & 1.12 & 1.20 \\
         &                   & 0.19 & 0.18 &0.074 & 1.00 & 2.15 \\
         &                   & 0.08 & 0.08 &0.023 & 1.00 & 3.48 \\
    Sb   &     10.40-10.80   & 0.32 & 0.28 &0.564 & 1.17 & 0.79 \\
         &                   & 0.42 & 0.41 &0.224 & 1.22 & 1.80 \\
         &                   & 0.22 & 0.24 &0.066 & 1.00 & 2.80 \\
         &                   & 0.05 & 0.07 &0.019 & 1.00 & 3.36 \\
    Sb   &     10.80-11.05   & 0.34 & 0.25 &0.535 & 1.28 & 0.92 \\
         &                   & 0.39 & 0.42 &0.151 & 1.36 & 2.49 \\
         &                   & 0.22 & 0.26 &0.069 & 1.00 & 3.22 \\
         &                   & 0.05 & 0.08 &0.023 & 1.00 & 3.53 \\
    Sb   &     11.05-11.90   & 0.23 & 0.17 &0.239 & 1.50 & 1.32 \\
         &                   & 0.39 & 0.40 &0.161 & 1.15 & 2.29 \\
         &                   & 0.30 & 0.33 &0.081 & 1.00 & 2.82 \\
         &                   & 0.07 & 0.10 &0.049 & 1.00 & 2.05 \\
         \hline
 Sc+Sd   &      8.67-10.00   & 0.20 & 0.23 &0.342 & 1.20 & 0.71 \\
         &                   & 0.40 & 0.37 &0.308 & 1.05 & 1.05 \\
         &                   & 0.24 & 0.24 &0.148 & 1.00 & 1.24 \\
         &                   & 0.17 & 0.16 &0.206 & 0.83 & 1.14 \\
 Sc+Sd   &     10.00-10.40   & 0.29 & 0.28 &0.415 & 0.96 & 0.76 \\
         &                   & 0.42 & 0.41 &0.374 & 1.00 & 1.01 \\
         &                   & 0.16 & 0.18 &0.075 & 1.00 & 1.67 \\
         &                   & 0.13 & 0.13 &0.080 & 0.83 & 1.71 \\
 Sc+Sd   &     10.40-10.80   & 0.37 & 0.31 &0.441 & 1.50 & 0.93 \\
         &                   & 0.44 & 0.42 &0.612 & 0.78 & 0.66 \\
         &                   & 0.15 & 0.20 &0.052 & 1.00 & 2.31 \\
         &                   & 0.04 & 0.07 &0.089 & 0.20 & 2.02 \\
 Sc+Sd   &     10.80-11.05   & 0.34 & 0.28 &0.289 & 1.35 & 1.01 \\
         &                   & 0.51 & 0.47 &0.123 & 1.16 & 1.15 \\
         &                   & 0.14 & 0.25 &0.042 & 1.00 & 1.96 \\
         &                   & 0.00 & 0.00 &0.010 & 0.70 & 5.00 \\

\hline
  \hline
 \end{tabular}
\end{table*}

\subsubsection{Sersic profile}
\label{SSS:sersic}

\begin{figure}
\centering\includegraphics[width=8.4cm]{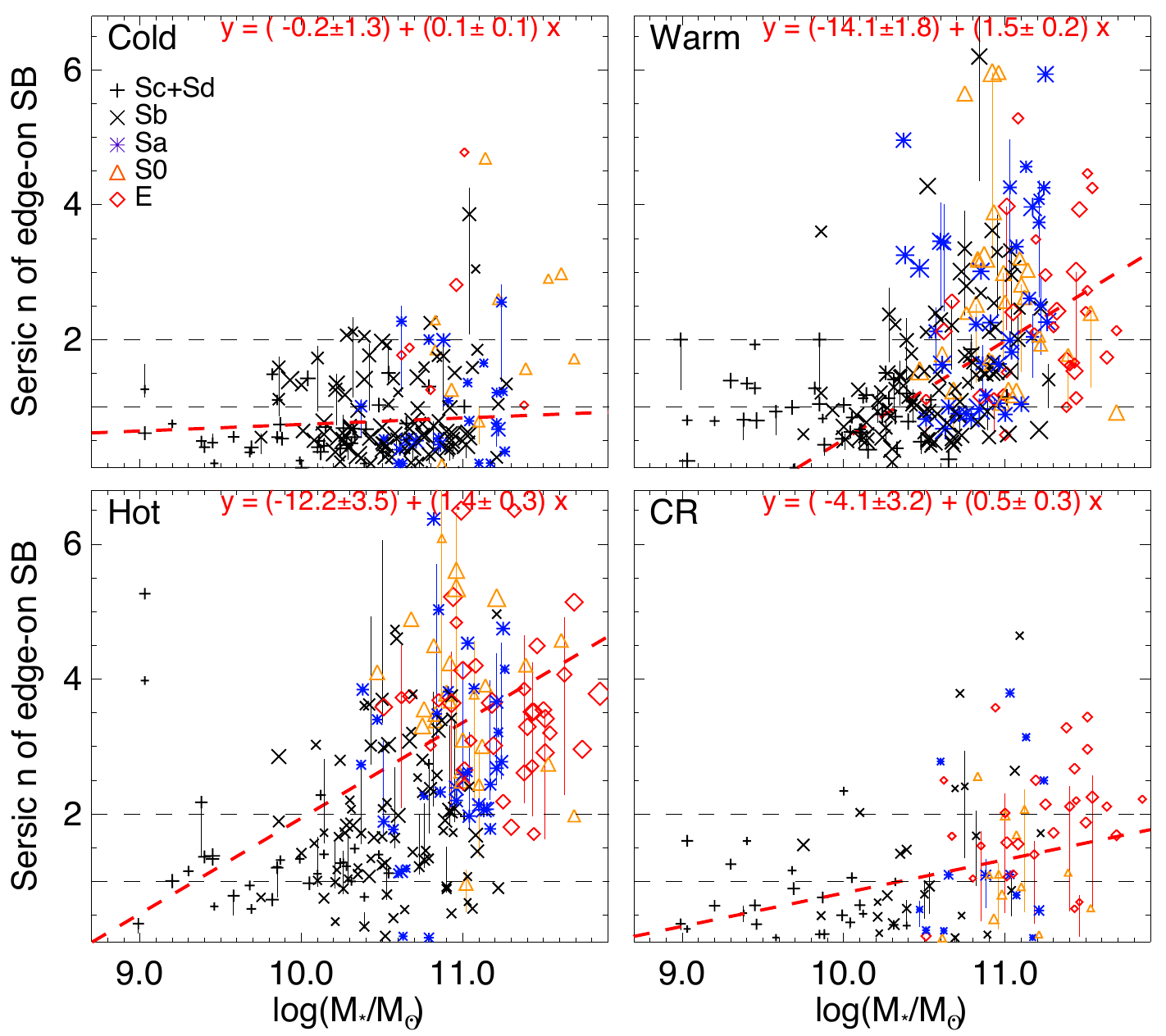}
\caption{The resulting Sersic index \textit{n} of cold, warm, hot, and CR components as a function of galaxy's total stellar mass $M_{*}$. Black pluses, black crosses, blue asterisks, yellow triangles, and red diamonds represent Sc/Sd, Sb, Sa, S0, and E galaxies, respectively, with symbol size indicating the orbital fraction. We only obtain a reasonable fit when one component has a relative large luminosity fraction $\gtrsim0.1$.  As a result, there are 164, 230, 207, and 102 points shown in the panels of cold, warm, hot and CR components, respectively. The short vertical lines indicate the $1\sigma$ uncertainties of the points including the systematic bias and systematic errors as described in appendix~\ref{S:error}, we have errors for all, but only shown that for randomly $1/5$ of the points.
The red dashed line represents the least-$\chi^2$ linear fit as the equation labeled in each panel. The two horizontal dashed lines represent $n=1$ and $n=2$. }
\label{fig:sersicn_i90}
\end{figure}

We fit a Sersic profile:
\begin{equation}
\Sigma(r) = \Sigma_{\mathrm{e}} 10^{-b_n((r/r_s)^{1/n}-1)}
\end{equation}
to the SB of each component, where $b_n = 0.868n-0.142$ \citep{Caon1993}, the effective surface density $\Sigma_{\mathrm{e}}$, effective radius $r_s$ and Sersic index \textit{n} are the three free parameters. 

The kinematic data of all these 250 galaxies extend to at least $1\,R_e$ and for 200 galaxies to $1.5\,R_e$. The orbit distribution, and therefore the SB profiles of each component are reliable within the data coverage. 
With reliable SB profiles extending to $\sim 1.5 R_e$, the parameters ($\Sigma_{\mathrm{e}}$, $r_s$, $n$) are not strongly constrained. In particular, $r_s$ and \textit{n} are degenerated, so that we put constraints of $r_s < 2R_e$, $n < 6.5$ for the cold and warm components, $r_s < R_e$, $n < 6.5$ for the hot and CR component. 

Per bin in stellar mass and Hubble type, we average the SB profiles of the cold, warm, hot and CR components.
The blue, orange, red, and black curves in Figure~\ref{fig:sb300} represent the best-fit Sersic profiles to these SB profiles with parameters listed in Table~\ref{tab:sb300}. 

We also fit the four components of each single galaxy separately. We consider the fit reasonable if it converges without hitting the boundaries of Sersic index \textit{n} and matches the data well as checked by eye. Reasonable fits are not always reached, especially for those components with luminosity fractions smaller than 0.1. We have obtained good fits for 164, 230, 207, 102 of the 250 galaxies for the cold, warm, hot and CR component, respectively.  
As shown in Figure~\ref{fig:maps5}, the variation of SB of each components within $1\,\sigma$ confidence level is small, which leads to a small statistical error of the Sersic index \textit{n} of $\sigma(n)/n \sim 10\%$. The systematic bias and error, however, are much larger as shown in Appendix~\ref{S:error}. In the following sections,  errors of all the parameters describing morphology are inferred from the systematic bias and errors listed in Table~\ref{tab:error}.

In Figure~\ref{fig:sersicn_i90}, we show the resulting Sersic index \textit{n} of cold, warm, hot, and CR components as a function of a galaxy's total stellar mass $M_{*}$. Black pluses, black crosses, blue asterisks, yellow triangles and red diamonds represent Sc+Sd, Sb, Sa, S0 and E galaxies, respectively.

\begin{table*}
\caption{Average structure parameters of the four orbital components. Sersic \textit{n} and scale radius $r_s$ in unit of galaxy's $R_e$ are from the Sersic fit to edge-on surface brightness radial profiles, $C$ is concentration, $q_{\mathrm{Re}}$ and $q_{\mathrm{Rmax}}$ are intrinsic flattening calculated within the half-light-radius $R_{\mathrm{e}}$ and the data coverage $R_{\mathrm{max}}$, respectively. For \textit{n} and $r_s$, the central value and error bar are the average and $1\sigma$ variation among these galaxies with reasonable Sersic fits; there are 164, 230, 207 and 102 for cold, warm, hot and CR, respectively.
Any components contributing a fraction larger than $0.05$ are included for calculating the concentration $C$ and flattening $q_{\mathrm{Re}}, \, q_{\mathrm{Rmax}}$; there are 232, 250, 248, 152 points for cold, warm, hot and CR components.  Similar structure parameters for each of the 250 galaxies are given in Table~\ref{tab:orbitf} in the appendix.}
\label{tab:param}\footnotesize
\begin{tabular}{*{6}{l}}
\hline
Components & $n$ &  $r_s/R_e$ & $C$ & $q_{\mathrm{Re}}$ & $q_{\mathrm{Rmax}}$ \\
\hline
Cold & $1.0\pm0.8$ & $1.3\pm0.5$ & $0.6\pm0.3$ & $0.37\pm0.15$ & $0.23\pm0.10$ \\
Warm & $2.0\pm1.5$ & $1.0\pm0.4$ & $1.1\pm0.5$ & $0.57\pm0.21$ & $0.44\pm0.22$ \\
Hot  & $2.7\pm1.6$ & $0.8\pm0.3$ & $1.5\pm0.5$ & $0.77\pm0.18$ & $0.67\pm0.22$ \\
CR   & $2.0\pm2.0$ & $0.7\pm0.4$ & $1.6\pm1.1$ & $0.57\pm0.23$ & $0.48\pm0.25$ \\
  \hline
 \end{tabular}
\end{table*}

The averaged effective radius $r_s$ and Sersic index \textit{n} of the four components are given in Table~\ref{tab:param}. The values are averaged from the 164, 230, 207, 102 points for cold, warm, hot and CR components, while the errors are the $1\,\sigma$ variations among these galaxies.
The cold components have similar Sersic index \textit{n} in almost all types of galaxies across the whole mass range, except in the most massive galaxies, in which the cold component is usually insignificant.
For the warm and hot components, \textit{n} increases with increasing stellar mass $M_*$. 
A best linear fit yields $n = (-14.1\pm1.8) + (1.5\pm0.2) \times \log10(M_*)$ for the warm component, with Pearson correlation coefficients (hereafter PCC) of 0.44 indicating a mild positive correlation. 
A best linear fit yields $n = (-12.2\pm3.5) + (1.4\pm0.3) \times \log10(M_*)$ for the hot component, with PCC of 0.41.
The CR component usually declines fast with radius, with \textit{n} comparable to that of the warm/hot components. 



\subsubsection{Concentration}
\label{SSS:concentration}
\begin{figure}
\centering\includegraphics[width=8.4cm]{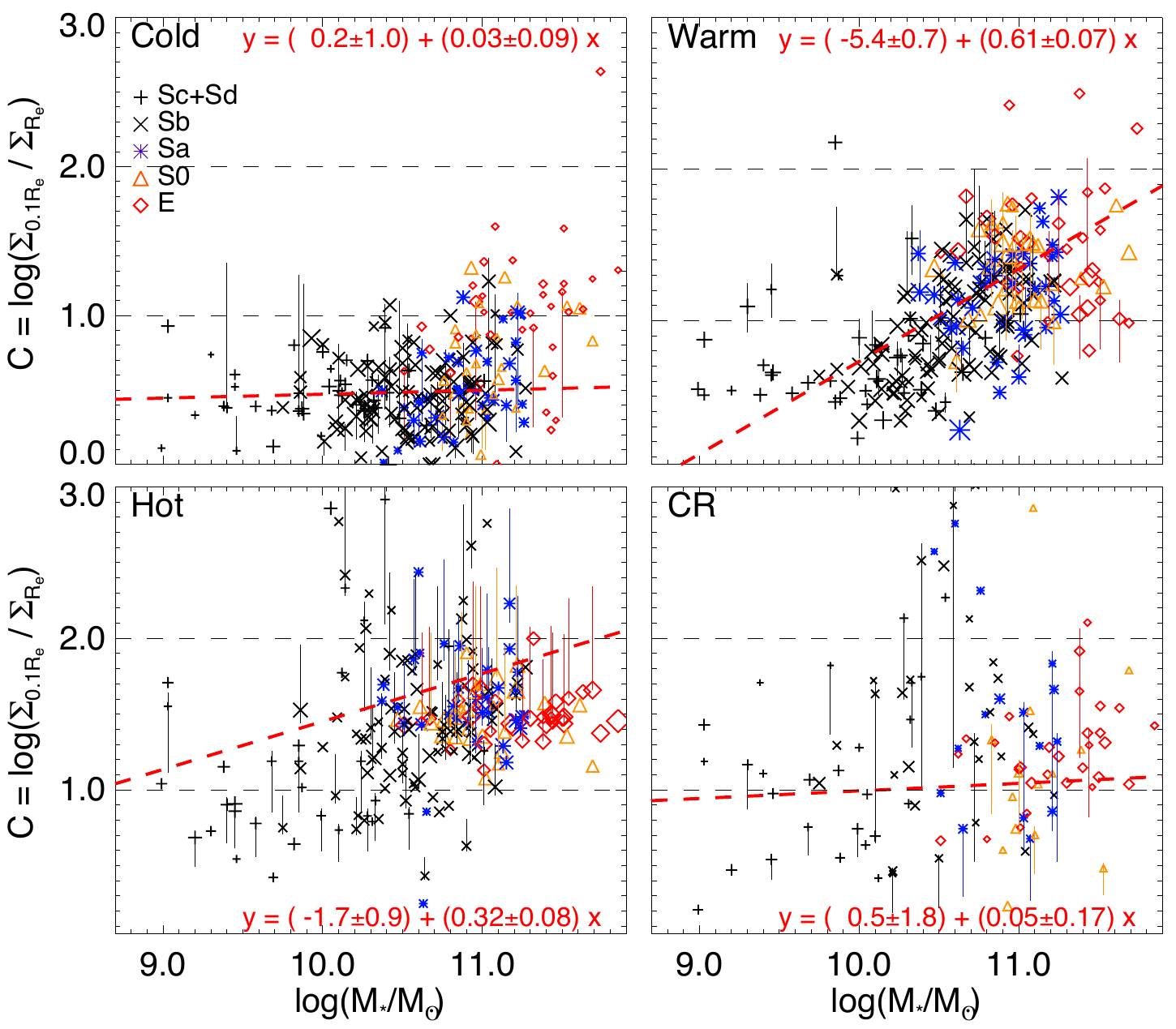}
\caption{Concentration as obtained in Equation~\ref{eq:C} of the four components as a function of total stellar mass $M_{*}$. The symbols are the same as in Fig~\ref{fig:sersicn_i90}. 
Any component which contributes a fraction larger than $0.05$ is included for calculating concentration $C$, as a result there are 232, 250, 248, 152 points for the cold, warm, hot and CR components. The short vertical lines indicate the $1\sigma$ uncertainties of the points including the systematic bias and systematic errors as described in appendix~\ref{S:error}, we have errors for all, but only shown that for randomly $1/5$ of the points. The red dashed line represents the fit as indicated by the equation in each panel. The two horizontal dashed lines represent $C=1$ and $C=2$, respectively.}  
\label{fig:concentration}
\end{figure}

Alternatively, the radial SB distribution can be quantified by the concentration:
\begin{equation}
\label{eq:C}
C = \log{(\Sigma_{0.1R_e}/\Sigma_{R_e})},
\end{equation}

where $\Sigma_{0.1R_e}$ and $\Sigma_{R_e}$ are the edge-on surface density in the center $|R-0.1R_e| < 0.1 R_e$ and around the effective radius $|R-R_e| < 0.1R_e$ respectively.
The concentration could be used as a model-independent alternative to the Sersic index \textit{n}. Any components contributing a fraction larger than $0.05$ are included for calculating concentration $C$; there are 232, 250, 248, 152 points for cold, warm, hot and CR components. 

The concentrations of the four components as function of total stellar mass are shown in Figure~\ref{fig:concentration} and averages are given in Table~\ref{tab:param}. 
For the cold component, concentration keeps being low. For the warm and hot components, there is a clear increase of concentration with total stellar mass. 
A best fit yields $C = (-5.4\pm0.7) + (0.61\pm0.07)\times \log10(M_*)$ for the warm component and $C = (-1.7\pm0.9)+ (0.32\pm0.08)\times \log10(M_*)$ for the hot component. The correlation is stronger for the warm component with a PCC of 0.54, for the hot component with a PCC of 0.25.
The concentration of CR component is similar to that of the warm and hot component. 

\subsection{Intrinsic flattening}
\label{SS:edge}
\begin{figure}
\centering\includegraphics[width=8.4cm]{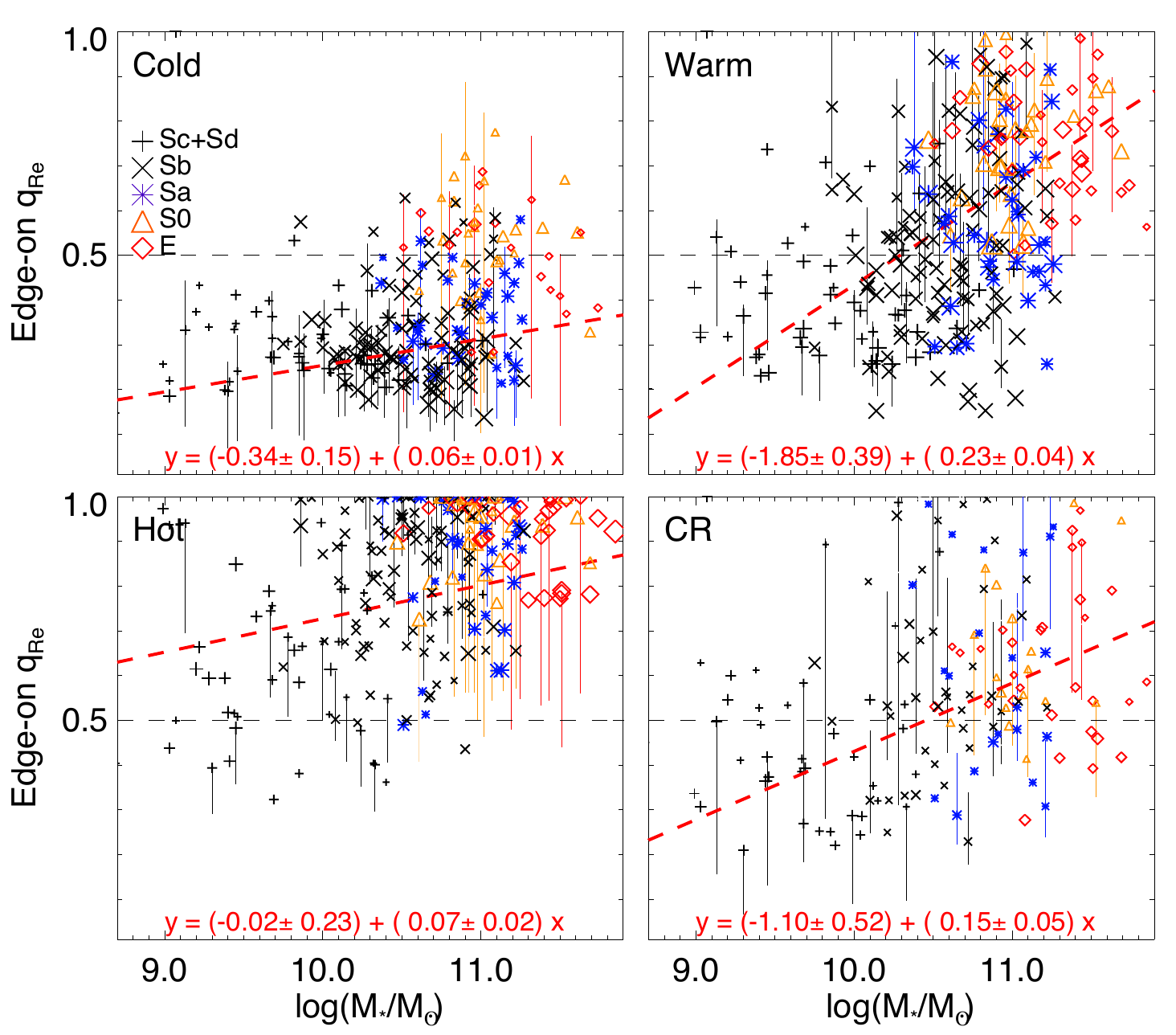}
\caption{The intrinsic flattening within the effective radius $q_{\mathrm{Re}}$ of the four components as a function of the galaxy's total stellar mass $M_*$. The symbols are the same as Fig~\ref{fig:sersicn_i90}. Any components contribute a fraction larger than $0.05$ are included for calculating intrinsic flattening; there are 232, 250, 248, 152 points for cold, warm, hot and CR components. The short vertical lines indicate the $1\sigma$ uncertainties of the points including the systematic bias and systematic errors as described in appendix~\ref{S:error}, we have errors for all, but only shown that for randomly $1/5$ of the points. The red dashed line represents the fit indicated by the equation in each panel. The horizontal dashed line represents $q_{\rm Re}= 0.5$.}
\label{fig:qq}
\end{figure}

We measure the average ellipticity $\epsilon$ from the edge-on SB maps.
Then the intrinsic flattening $q = b/a = 1-\epsilon$, where a and b are the long and short axis of each component \footnote{Note here for measuring $\epsilon$, the long and short axises are defined for each component; axis misalignment is allowed for different components in the same galaxy. The definition of intrinsic flattening $q$ in this way especially for the hot component is different from what was used in \citet{Zhu2018b}, where a and b are always aligned with major and minor axis of the galaxy.} 

The data coverage, and thus the reliable regions for the SB of each component, are not the same for all galaxies. Here we calculate $q_{\rm Re}$ using only the regions within $R_e$, and $q_{\rm Rmax}$ using the regions within data coverage $R_{\rm max}$ with average $R_{\rm max}/R_e \sim 2$ for the sample. The former is a uniform value which can be used to compare with simulations, while the latter is our best comparison with the observed flattening for disks. 

The flattenings $q_{\mathrm{Re}}$ of the four components as function of total stellar mass $M_*$ are shown in Figure~\ref{fig:qq} and average values are given in Table~\ref{tab:param}.
All four components have clear trends of becoming thicker/rounder in more massive galaxies.
The cold component keeps thin $q_{\rm Re} \sim 0.3$ for most of low-mass late-type galaxies, and becomes thicker with $q_{\rm Re} \sim 0.5$ in massive early-type galaxies, with a best-linear fit of $q_{\rm Re} = (-0.34\pm0.15) + (0.06\pm0.01)\times \log10(M_*)$.
The warm component is flat with $q_{\rm Re} \sim 0.4$ in late-type galaxies, and becomes round with $q_{\rm Re} \sim 0.8$ in massive early-type galaxies, with a best-linear fit of $q_{\rm Re} = (-1.85\pm0.39) + (0.23\pm0.04)\times \log10(M_*)$.
The hot component is near spherical with $q_{\rm Re} \sim 0.8$ in most galaxies, except in some low-mass late-type galaxies with flattening as low as $q_{\rm Re} \sim 0.5$, with a best-linear fit of $q_{\rm Re} = (-0.02\pm0.23) + (0.07\pm0.02)\times \log10(M_*)$. The CR component follows the warm/hot component. 

\begin{figure}
\centering\includegraphics[width=8.4cm]{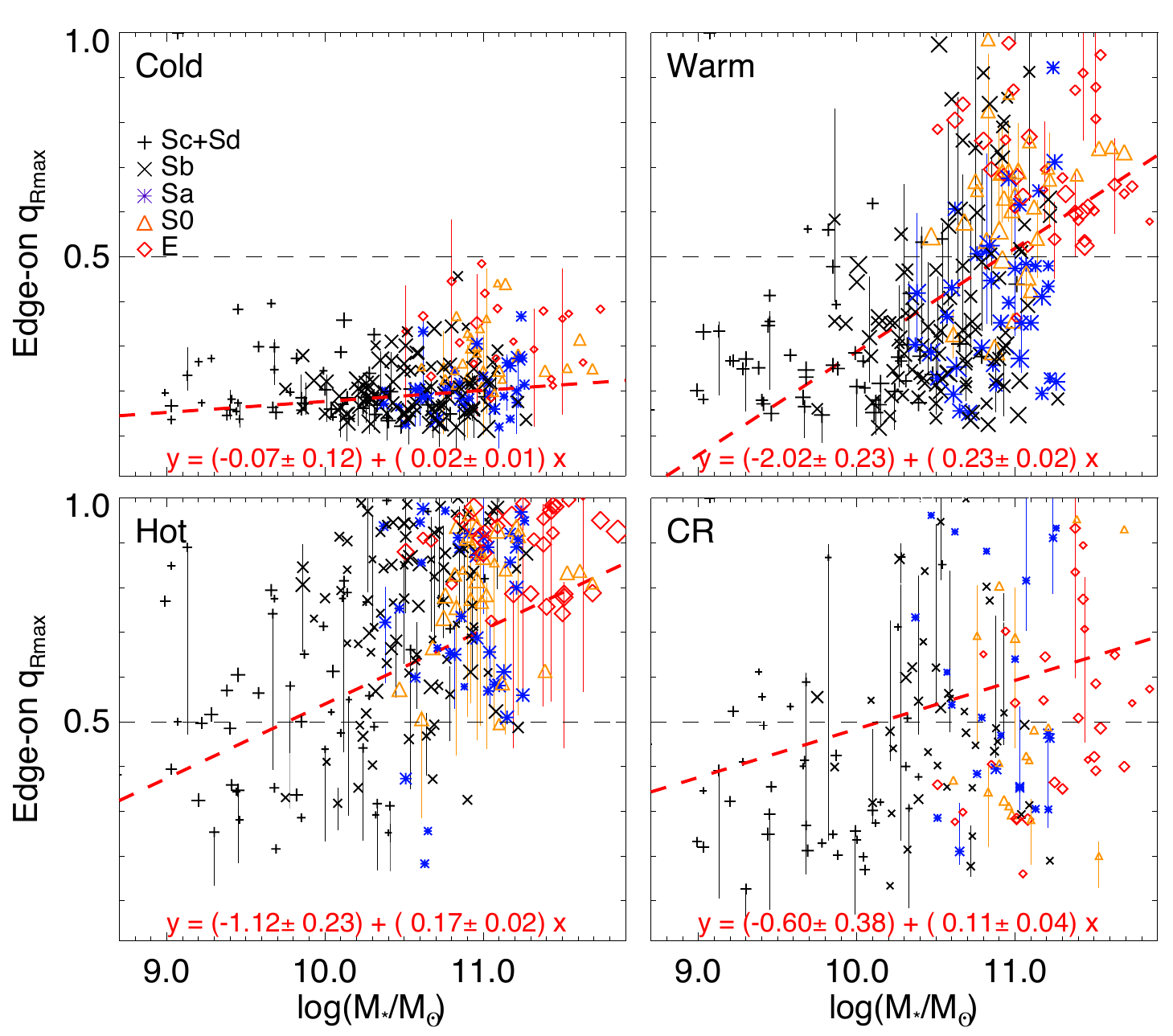}
\caption{The intrinsic flattening $q_{\mathrm{Rmax}}$, measured within the CALIFA data coverage $R_{\rm max}$, of the four components as a function of the galaxy's total stellar mass $M_*$. The symbols are the same as Figure~\ref{fig:qq}. The red dashed line represents the fit indicated by the equation in each panel. The horizontal dashed line represents $q_{\rm Rmax}= 0.5$.}
\label{fig:qqmax}
\end{figure}

Similarly, the flattenings $q_{\mathrm{Rmax}}$ of the four components as function of total stellar mass $M_*$ are shown in Figure~\ref{fig:qqmax}. 

The variety of $q$ indicates that it is not always true for flattened (disk-like) structures to be rotating, or round (bulge-like) ones to be non-rotating. There are flat, disk-like hot components in low-mass late-type galaxies and also round, bulge-like warm component in high mass early-type galaxies. However, at fixed mass or in the same galaxy, flatter components generally have stronger rotation than rounder components. 

\subsection{Radial distribution vs. intrinsic flattening}
\label{S:nvsq}
\begin{figure}
\centering\includegraphics[width=8.4cm]{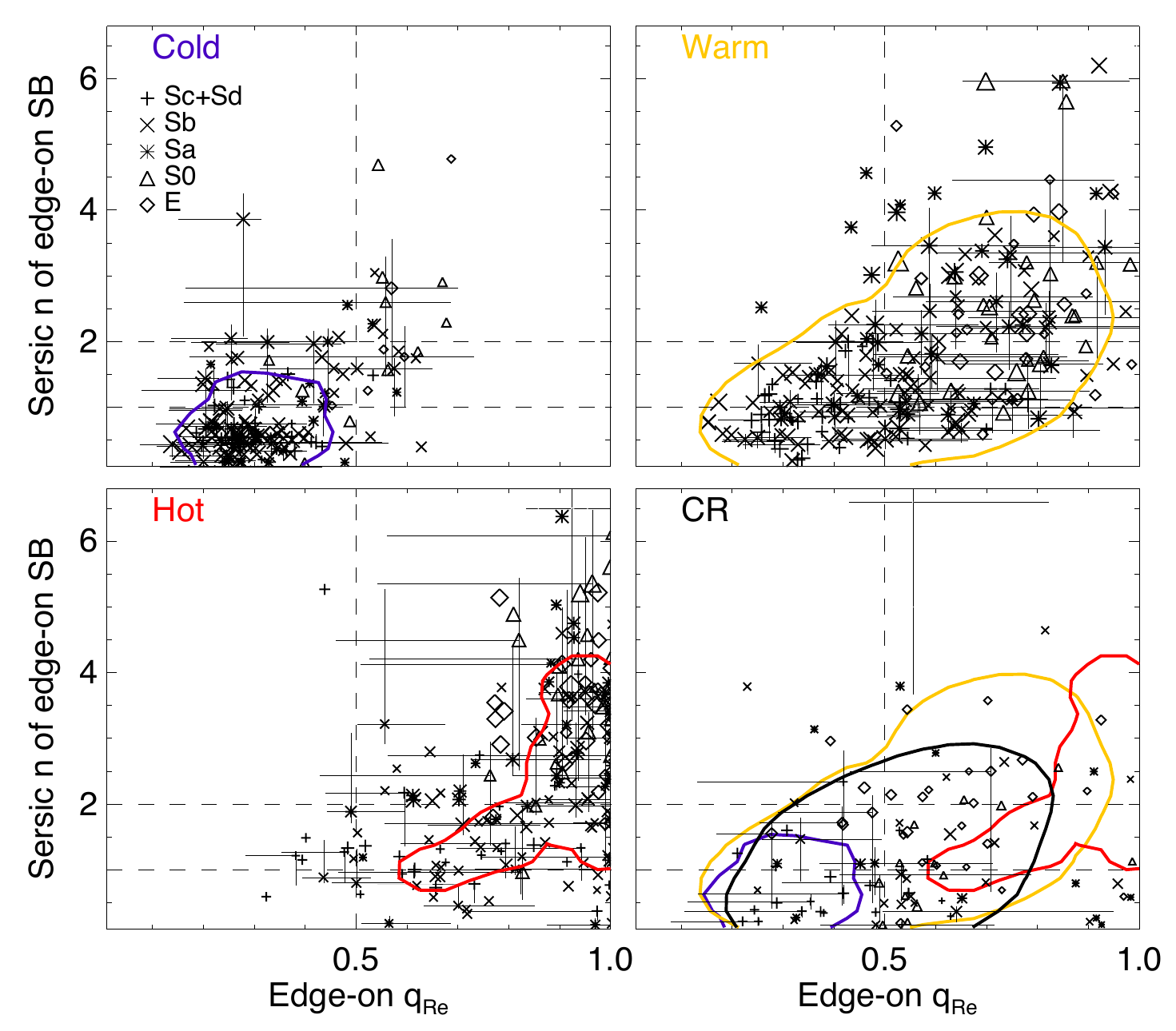}
\caption{Sersic index \textit{n} vs. the intrinsic flattening $q_{\mathrm{Re}}$ of the four components. The four panels represent cold, warm, hot and CR components, each symbol is one galaxy. The short horizontal and short vertical lines indicate the $1\,\sigma$ uncertainties of $q_{\rm Re}$ and $n$, including the systematic bias and systematic errors as shown in the Appendix~\ref{S:error}, we have errors for all, but only shown for randomly  1/5 of the points. Similar to Figure~\ref{fig:sersicn_i90}, one component is only shown when a reasonable Sersic fitting is obtained. The blue, yellow, red and black contours enclose $\sim 80\%$ of the points for cold, warm, hot and CR components in the diagram. The vertical dashed represents $q_{\rm Re} = 0.5$, while the two horizontal dashed lines are $n=1$ and $n=2$, respectively. In the CR panel, we overplot the contours of all four components for comparison. In general, one component has larger Sersic \textit{n} when it is rounder. The cold and hot components are well separated in this diagram. Warm component has overlaps with cold and hot components at the two ends.}
\label{fig:sersicn_qqi90}
\end{figure}

In Figure~\ref{fig:sersicn_qqi90}, we show the occupancy of the four components in the Sersic \textit{n} versus $q_{\mathrm{Re}}$ diagram. 
The blue, yellow, red and black contours enclose $\sim 80\%$ of the points for cold, warm, hot and CR components in the diagram. The four components show similar trends of increasing Sersic \textit{n} with increasing $q_{\mathrm{Re}}$, thus generally, any component rounder tends to be more radially concentrated (with larger \textit{n}). 

Systematically, $q_{\mathrm{Re}}$ and \textit{n} increase from cold to hot components, but there are overlaps as shown in the bottom-right panel. The warm components largely overlap with cold components at the side with lower $q_{\mathrm{Re}}$ and \textit{n}, and overlap with hot components at the other side. 
Cold and hot are well separated in the diagram with little overlap. 
The CR component covers similar regions as the warm component. 

\begin{figure}
\centering\includegraphics[width=8.4cm]{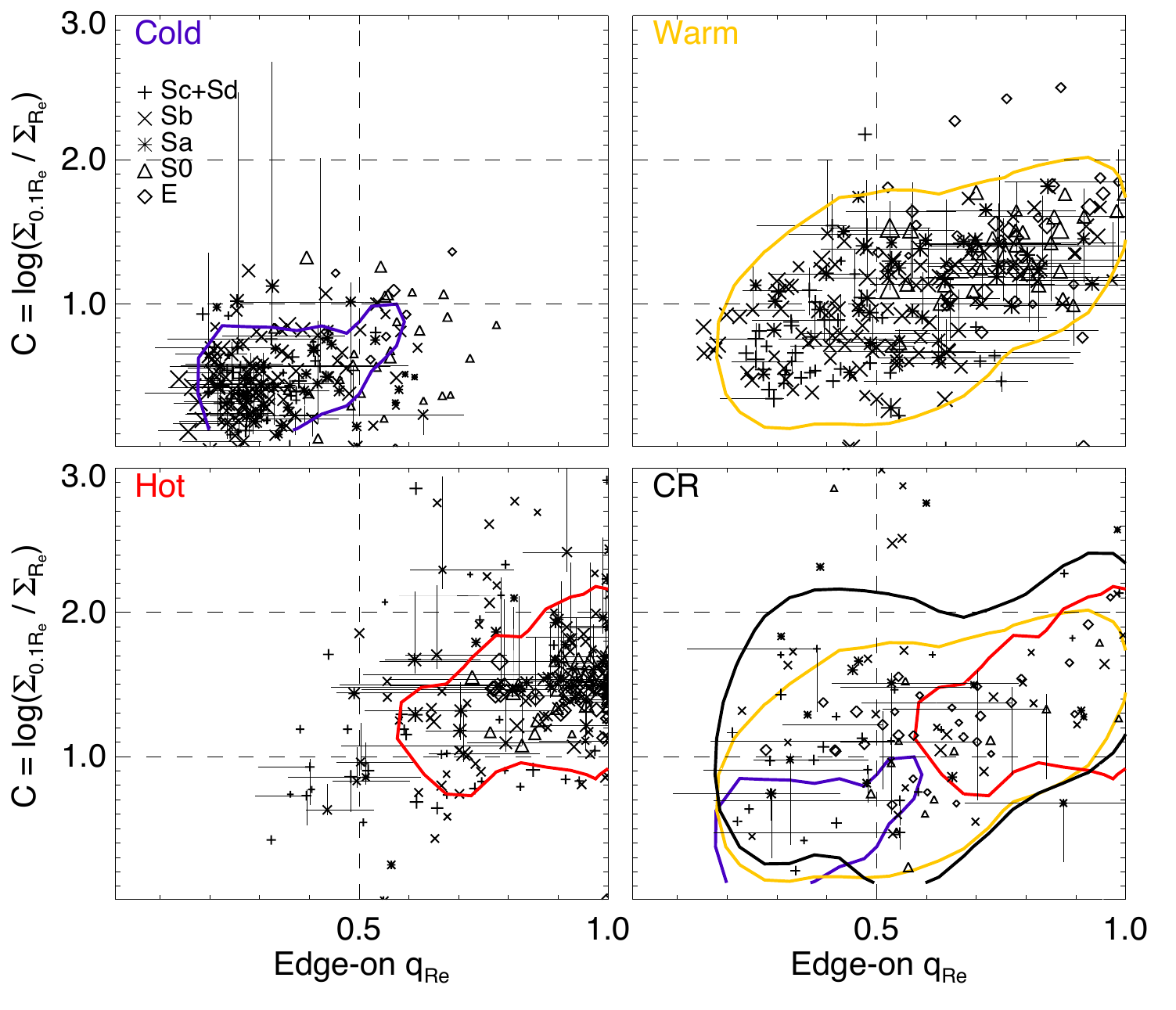}
\caption{Concentration $C$ vs. the intrinsic flattening $q_{\mathrm{Re}}$. The short horizontal and short vertical lines indicate the $1\,\sigma$ uncertainties of $q_{\rm Re}$ and $C$, including the systematic bias and systematic errors as shown in the Appendix~\ref{S:error}, we have errors for all, but only shown for randomly 1/5 of the points. The blue, yellow, red and black contours enclose $\sim 80\%$ of the points for cold, warm, hot and CR components in the diagram. The vertical dashed represents $q_{\rm Re} = 0.5$, while the two horizontal dashed lines are $C=1$ and $C=2$, respectively. In the CR panel, we overplot the contours of all four components for comparison. In general, one component is more concentrated when it is rounder. The cold and hot components are well separated on this diagram, while the warm component overlaps with the cold and hot components at both ends.}
\label{fig:C_qqi90}
\end{figure}

In Figure~\ref{fig:C_qqi90}, we show a similar diagram of $C$ vs. $q_{\mathrm{Re}}$, using concentration $C$ instead of Sersic \textit{n}. The four components show similar trends of increasing concentration $C$ with increasing $q_{\mathrm{Re}}$. 
Cold and hot components are well separated, while the warm components are still overlapping  with the cold and hot components at both ends. 

\section{Comparison with results from photometric studies}
\label{S:observation}
\subsection{Bulge vs. disk fractions}
\label{SS:photmetric}

\begin{figure}
\centering\includegraphics[width=8.4cm]{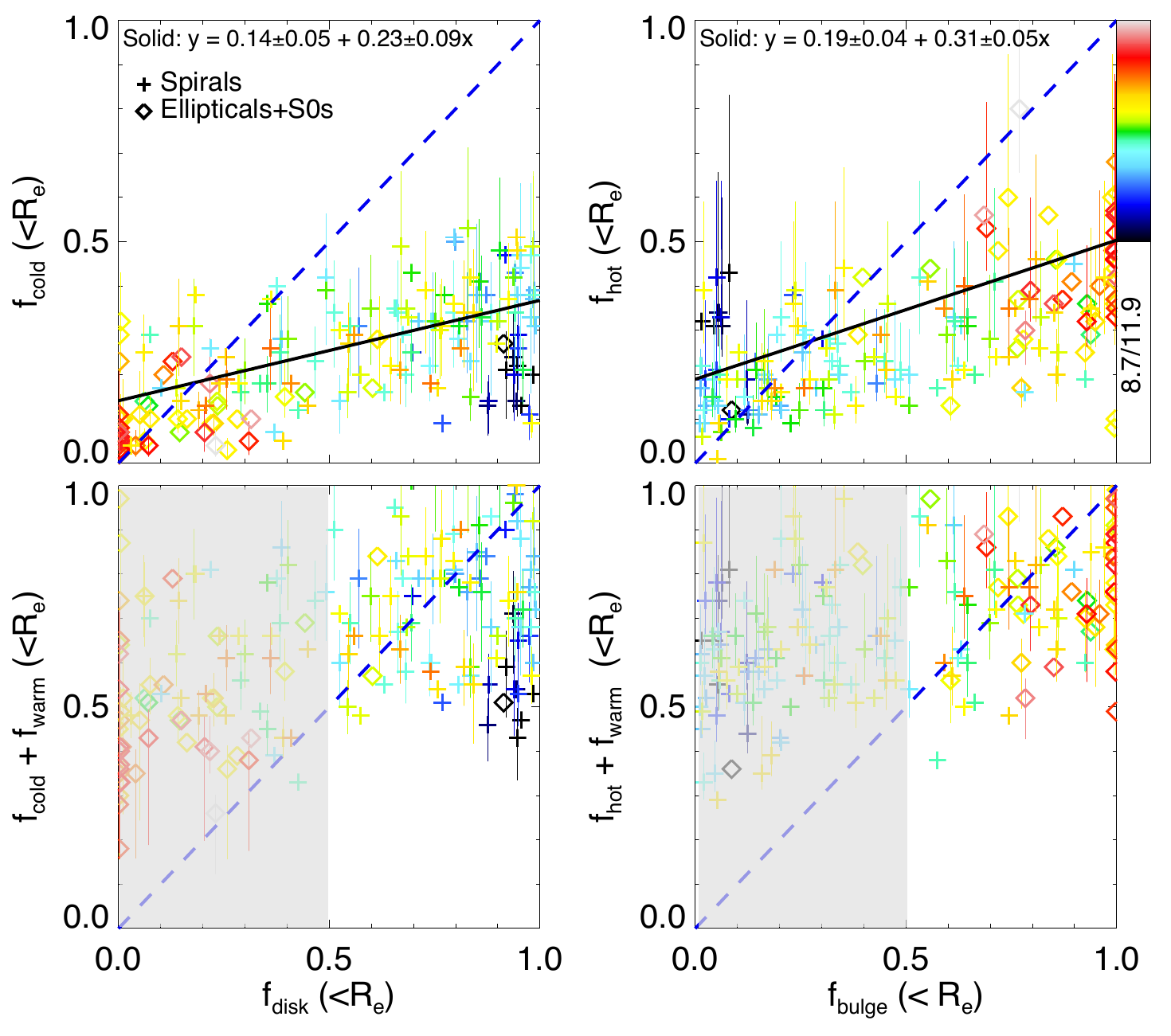}
\caption{Orbital fractions compared with the photometric bulge/disk fraction; $f_{\mathrm{cold}}$ vs. $f_{\mathrm{disk}}$ (top-left), $f_{\mathrm{hot}}$ vs. $f_{\mathrm{bulge}}$ (top-right), $f_{\mathrm{cold}}+f_{\mathrm{warm}}$ vs. $f_{\mathrm{disk}}$ (bottom-left), and $f_{\mathrm{hot}} + f_{\mathrm{warm}}$ vs. $f_{\mathrm{bulge}}$ (bottom-right). Each plus represents a late-type galaxy and a diamond represents an early-type galaxy, colored with their stellar masses in $\log10(M_*/M_{\odot})$ as indicated by the color bar. The short vertical lines indicate the $1\,\sigma$ uncertainties, including the systematic bias and systematic errors as described in Section~\ref{S:sample}, we have errors for all, but only shown that for randomly $1/2$ of the points. The blue dashed line represents $y = x$, the solid black line illustrates the best linear fits as indicated by the equations. The very low-mass galaxies with $M_* <10^{10} \, M_{\odot}$ (dark blue points) are outliers of these relations.
In low mass late-type galaxies, cold orbit fractions are smaller than their disk fractions, which the cold + warm orbit fractions match; in high-mass early-type galaxies, hot orbit fractions are smaller than their bulge fractions, which the hot + warm orbit fractions toward match.}
\label{fig:photometric}
\end{figure}

A galaxy's morphology is commonly fitted by a combination of a Sersic bulge and an exponential disk. From late-type to early-type galaxies, the contribution of bulge increases and that of disk decreases. 

Of the 250 galaxies, 190 galaxies have photometric decompositions published in \cite{Mendez-Abreu2017}. 
Our orbit fractions are only within $1\,R_e$ of the galaxies. To have a fair comparison with the photometric decomposition results, we also derive the bulge and disk fractions within $1\,R_e$ from the Sersic bulge and exponential disk in \cite{Mendez-Abreu2018}. As shown in the top two panels of Figure~\ref{fig:photometric}, the cold component fraction $f_{\rm cold}$ is positively correlated with disk fraction $f_{\rm disk}$ and the hot component fraction $f_{\rm hot}$ is positively correlated with bulge fraction $f_{\rm bulge}$. 
Linear fits to the points in Figure~\ref{fig:photometric} yield $f_{\mathrm{cold}} = (0.14\pm0.05) + (0.23\pm0.09) f_{\mathrm{disk}}$ with PCC of 0.73 and $f_{\mathrm{hot}} = (0.19\pm0.04) + (0.31\pm0.05) f_{\mathrm{bulge}}$ with PCC of 0.65.
The very low-mass galaxies with $M_* <10^{10} \, M_{\odot}$ are outliers of these relations and we have excluded them in the fits.
Note that the results of photometric decomposition could be model-dependent, here we just use one published photometric decomposition to illustrate the relation. 

In the previous sections, we have shown that the cold component is generally disk-like with radial profile close to exponential with Sersic index $n \lesssim 1$.
The hot and CR components are spheroidal with radial profiles consistent with Sersic profile with index $n \gtrsim 2$ in most galaxies.

There is not a clear correlation of the warm orbit fractions with any of the morphological components (i.e. disk, bulge or bar).
The warm component could contribute to either disk, or bulge, or both.
For most low-mass late-type galaxies, the cold orbit fractions are smaller than their photometric disk fractions, because a significant fraction is own to the warm component, which is flat and with low Sersic \textit{n} in low-mass galaxies. In the bottom-left panel of Figure~\ref{fig:photometric}, we see that adding up cold and warm components ($f_{\rm cold} + f_{\rm warm}$) matches the disk fractions $f_{\rm disk}$ for these low-mass late-type galaxies.
For the high-mass early-type galaxies with low disk fractions, the hot orbit fractions $f_{\rm hot}$ are smaller than their photometric bulge fractions $f_{\rm bulge}$. Besides the CR component also the warm component could significantly contribute to the bulge fraction as they become round and concentrated with high Sersic \textit{n}. ($f_{\rm how} + f_{\rm warm}$) tends to match $f_{\rm bulge}$ in the high-mass early-type galaxies as shown in the bottom-right panel of Figure~\ref{fig:photometric}.

The luminosity fraction of photometrical-bulge/disk as a function of total stellar mass show similar trend as that of our cold/hot orbit component fraction. (e.g., \citealt{Mendez-Abreu2017}). The correlation of $f_{\rm cold}$ versus $f_{\rm disk}$ and $f_{\rm hot}$ versus $f_{\rm bulge}$ are consistent with this result.
\subsection{Sersic index of bulge and disk}
\label{SS:bulge}
\begin{figure}
\centering\includegraphics[width=7cm]{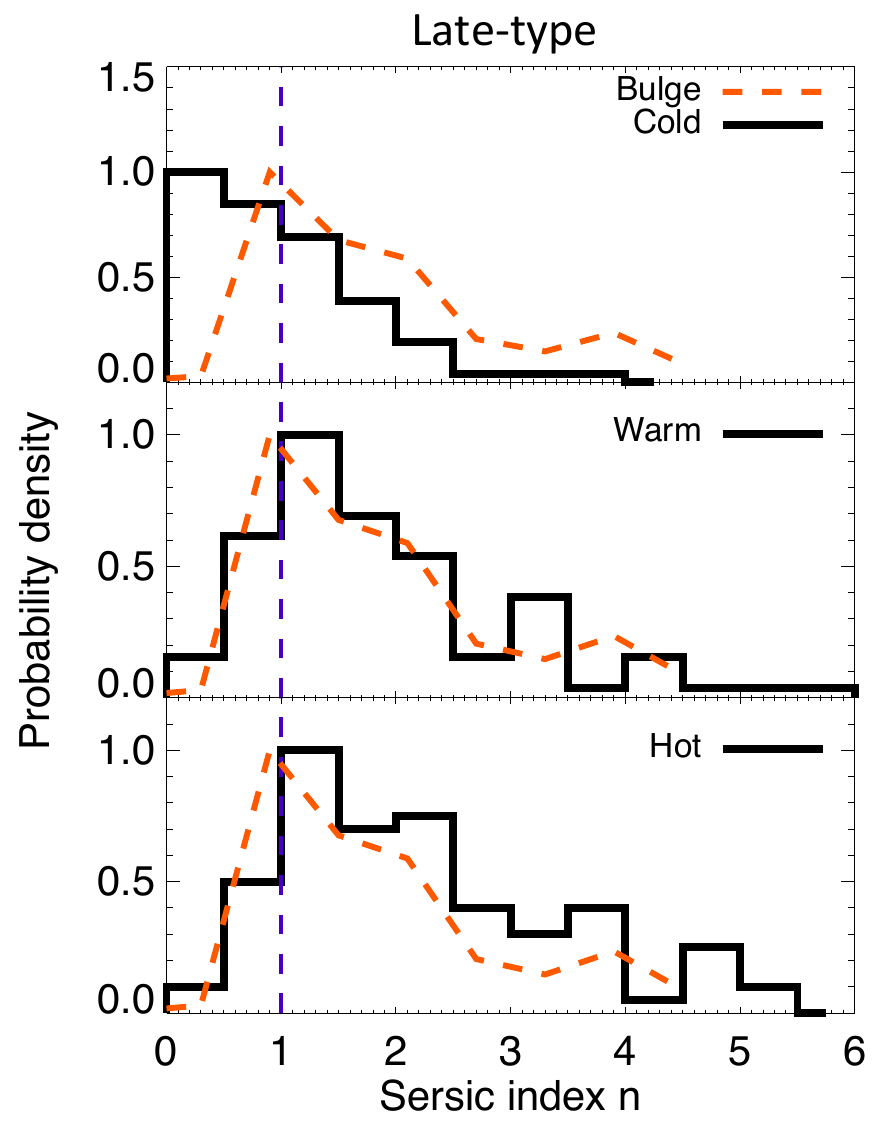}
\caption{Sersic index \textit{n} of the orbital components compared with those of the photometric bulges, for late-type galaxies only. The orange dashed curve is \textit{n} of bulges from \citet{Mendez-Abreu2017}, the black histogram are \textit{n} distribution of cold, warm, and hot components from top to bottom. The vertical dashed line is $n=1$, which is fixed for an exponential disk. }
\label{fig:bulge_spi}
\end{figure}

\begin{figure}
\centering\includegraphics[width=7cm]{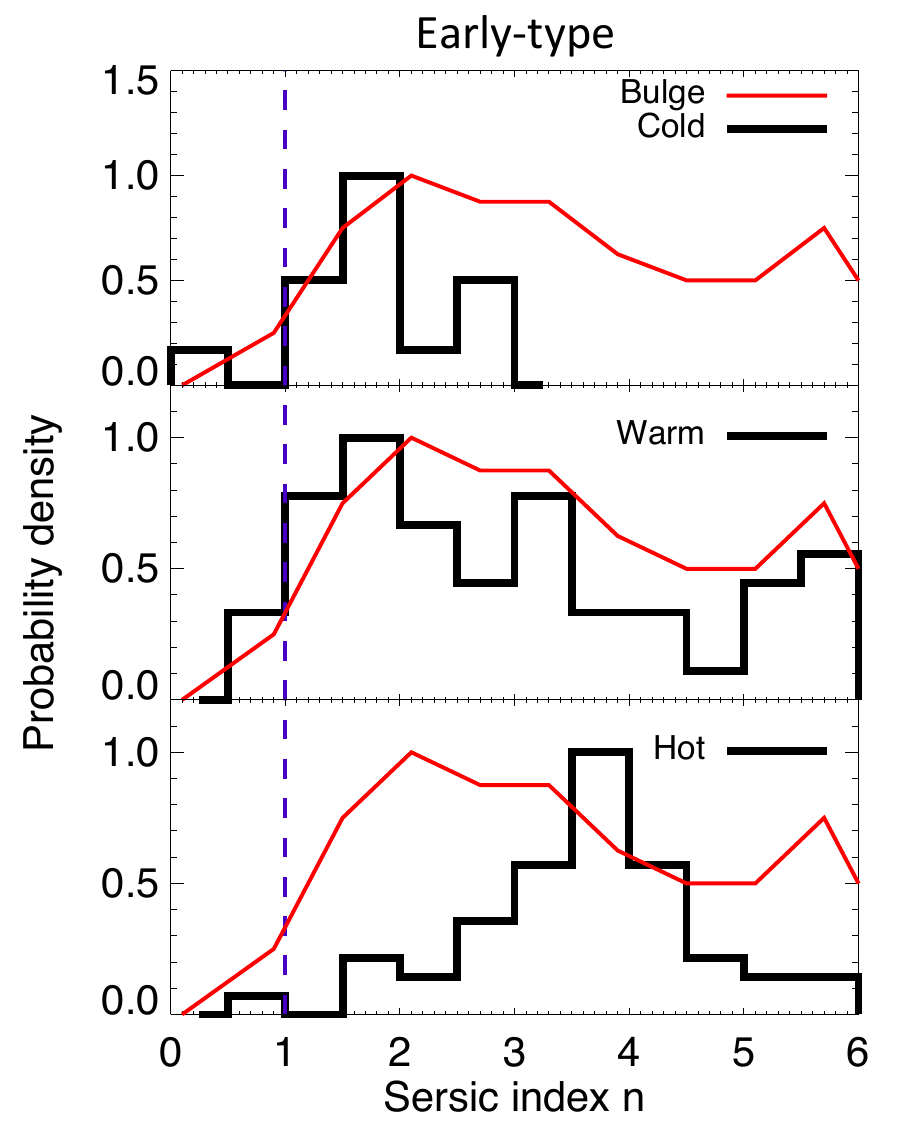}
\caption{Sersic index \textit{n} of the orbital components compared with those of the photometric bulges, but for early-type galaxies. The red solid curve is \textit{n} of bulges from \citet{Mendez-Abreu2017}, the black histogram are \textit{n} distribution of cold, warm, and hot components from top to bottom. The vertical dashed line is $n=1$, which is fixed for an exponential disk. }
\label{fig:bulge_ellp}
\end{figure}

Here we compare the Sersic index \textit{n} of the orbital component with that of the photometric-decomposed bulges and disks from \cite{Mendez-Abreu2017}. The galaxies are split into late-type galaxies in Figure~\ref{fig:bulge_spi} and early-type galaxies in Figure~\ref{fig:bulge_ellp}.

A photometrically-decomposed exponential disk has a fixed $n=1$. 
For spiral galaxies with usually a significant disk, the cold component has an average \textit{n} of 1, while there are a large fraction of cold components with $n<1$, consistent with the scenario that part of the disk is own to warm orbits. The warm and hot components have wide distributions of \textit{n} with similar centroids, generally consistent with that of photometric bulge. It is hard to tell either warm or hot orbits should contribute more to bulges from their \textit{n} distributions.
 
The early type galaxies do not have significant disks as late-type galaxies. The Sersic index \textit{n} of cold component is larger than $n=1$. \textit{n} increase from cold to warm and hot components. The photometric bulge has a wide distribution of \textit{n}. The cold component matches the lower end of the bulge, the hot component matches toward the higher end of bulge, while the warm component has a wide distribution of \textit{n} and touches the two ends of \textit{n} distribution of bulge. 
This is consistent with that bulges of massive early-type galaxies are not made only of hot orbits. 

The wide distribution of \textit{n} for different orbital components, as well as the general overlapping with photometric bulge implies that a Sersic index separation is not a good separation for non-rotating and rotating bulges.
This is consistent with the recent results from both observations \citep{Mendez-Abreu2018} and simulations \citep{Obreja2018b}. 
However, we would like to emphasis that, at a fixed stellar mass range, it is still true that bulges with higher Sersic index could have systematically lower rotation, knowing that hot components have systematically higher Sersic index \textit{n} than warm components as a function of stellar mass.

\subsection{Intrinsic flattening of bulge and disk}
\label{SS:intrinsic_flattening}
The flattening of bulges and disks directly from the photometric decomposition are the projected values, with the information diluted by galaxies' inclination angles. Thus we do not show the comparison of the projected flattening of our orbital components to those of bulges and disks from \citet{Mendez-Abreu2017}.
Instead, we compare statistically the intrinsic flattening of orbital components to those of disk and bulge from other studies.

\begin{figure}
\centering\includegraphics[width=7cm]{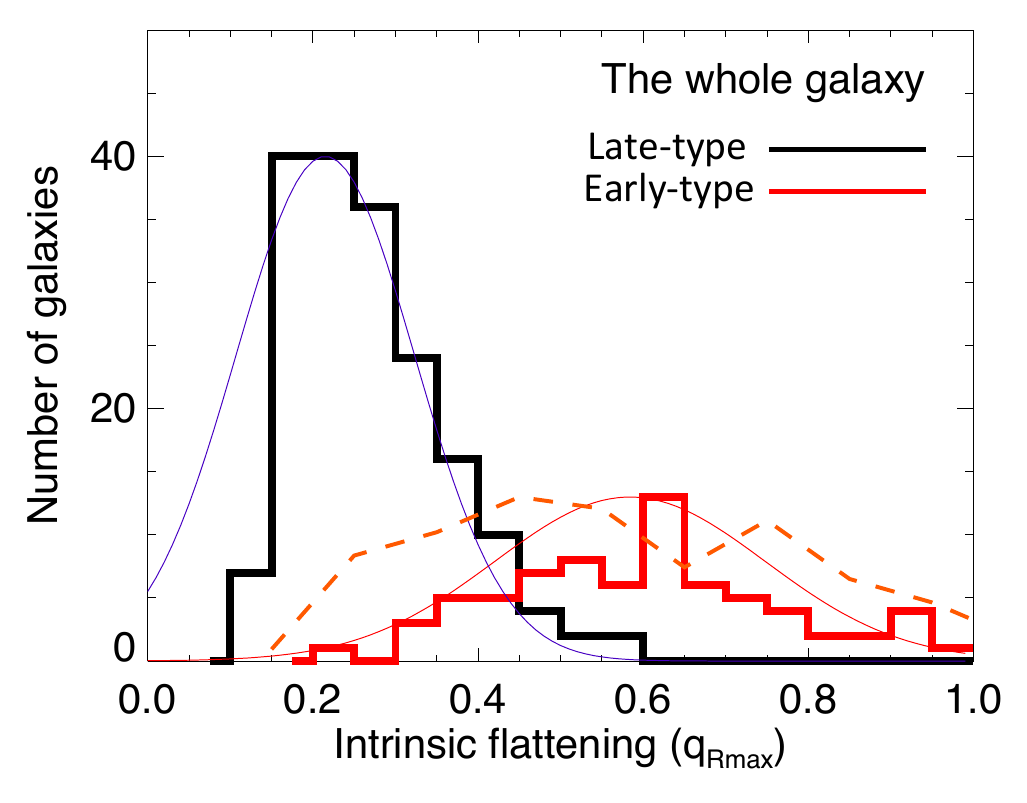}
\caption{Intrinsic flattening of galaxies as a whole. $q_{\mathrm{Rmax}}$ measured from our edge-on model for late-type galaxies in black solid histogram and early-type galaxies in black. The blue Gaussian profile with $(q_0, \sigma_q)=(0.215, 0.108)$ and red with $(q_0, \sigma_q)=(0.584, 0.164)$ are the distribution of intrinsic flattening of late-type and early-type galaxies inferred from large statistical sample of SDSS images \citep{Rodrguez2013}. The orange dashed curve represents the distribution of intrinsic flattening of bulges in CALIFA galaxies from \citet{Costantin2018}. The curves are normalized to match the height of the histograms. The intrinsic flattening of late-type and early-type galaxies we obtained are consistent with those from \citet{Rodrguez2013}, except that we are lacking of extremely flat disks ($q_{\mathrm{Rmax}} \lesssim 0.1$), which could partly caused by the limited spatial coverage of CALIFA data.}
\label{fig:q_ellp}
\end{figure}

We measure the intrinsic flattening of the whole galaxy within the data coverage $R_{\mathrm{max}}$, similarly to the way we measure that for each orbital components (see Section~\ref{SS:edge}). 
The galaxy's intrinsic flattening ($q_{\mathrm{Rmax}}$) distributions are shown in Figure~\ref{fig:q_ellp}, late-type galaxies in black solid histogram and early-type galaxies in red. We get an average $q_{\mathrm{Rmax}}$ of $0.275\pm0.106$ for spiral galaxies and $q_{\mathrm{Rmax}}$ of $0.697\pm0.168$ for elliptical galaxies.

The intrinsic flattening of late-type galaxies and early-type galaxies have been studied using large statistical samples of galaxy images (e.g., \citealt{Rodrguez2013}, which yield Gaussian distributions of $(q_0, \sigma_q)=(0.215, 0.108)$ (blue) and $(q_0, \sigma_q)=(0.584, 0.164)$ (red). 
The intrinsic shape of embedded central bulges are hard to measure. However, \citet{Costantin2018} measured the shape of bulges in 83 CALIFA galaxies using only photometric information, representing in the orange dashed curve in Figure~\ref{fig:q_ellp}.  

Our $q_{\mathrm{Rmax}}$ distributions generally match those based on SDSS images for late-type and early-type galaxies, respectively, with both similar central values and scatters. We are lacking extremely flat disks $q_{\mathrm{Rmax}} \lesssim 0.1$, which could be caused by the limited spatial coverage of the CALIFA data.

\begin{figure}
\centering\includegraphics[width=7.2cm]{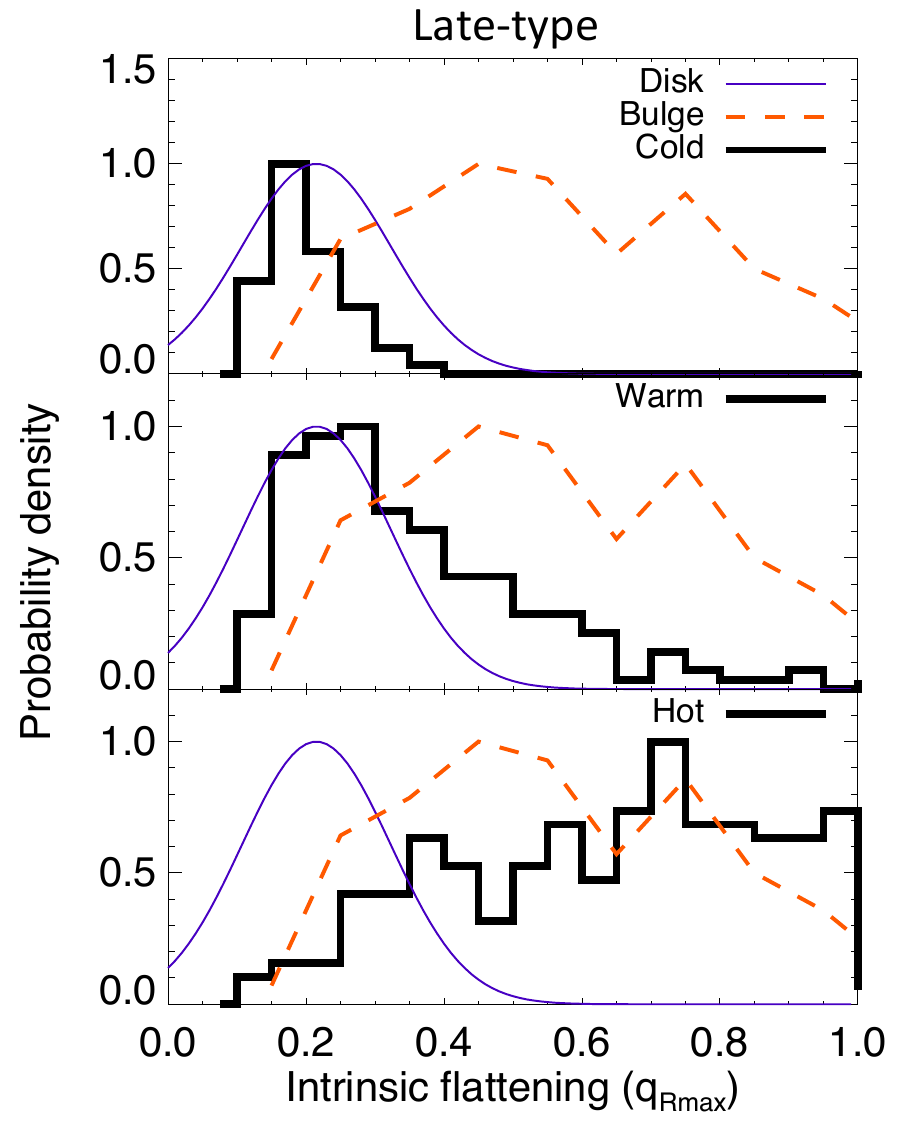}
\caption{The intrinsic flattening of the orbital cold, warm, hot components compared with those of photometric disks and bulges.
The black histograms are $q_{\mathrm{Rmax}}$ distribution of cold, warm, hot components, for late-type galaxies only, from top to bottom. The intrinsic flattening distribution of late-type galaxies (representative of disks) from \citet{Rodrguez2013} (blue solid) and bulges from \citep{Costantin2018} (orange dashed) are over plotted, the height of maximum being normalized to unity. It is consistent with the scenario of disk = cold + (warm), while bulge = hot + (warm) + (CR).}
\label{fig:q_spirals}
\end{figure}


We then focus on spiral galaxies, they have disks and bulges. In Figure~\ref{fig:q_spirals}, we show the flattening $q_{\mathrm{Rmax}}$ distribution of cold, warm and hot components as the thick histograms from top to bottom. The flattening distribution of late-type galaxies (representative of disks) from \citet{Rodrguez2013} (blue solid) and bulges from \citet{Costantin2018} (orange dashed) are over plotted.  

The peak of the cold component is $q_{\mathrm{Rmax}} \sim 0.17$, smaller than 0.215 of the disks. 
Thus, the cold components are generally thinner than photometric disks. Or in other words, disks are not only made of cold orbits. The $q_{\mathrm{Rmax}}$ distribution of the warm component peaks at $q_{\mathrm{Rmax}} \sim 0.25$, slightly higher than the 0.215 of disks. Warm orbits are consistent to be another main ingredient of disks. 

Hot orbits cover a large range of flattening, similar to the bulges. However, photometric bulges are not only made of hot orbits as their intrinsic flattening shows that some bulges are flatter than the hot component. The $q_{\mathrm{Rmax}}$ distribution of warm component has a long tail towards larger values overlapping with bulges, which is consistent with warm orbits also contributing significantly to bulges.

\section{Morphology versus kinematics}
\label{S:k-m}
\subsection{Random motion direction}
\label{SS:orbit}
\begin{figure}
\centering\includegraphics[width=7.5cm]{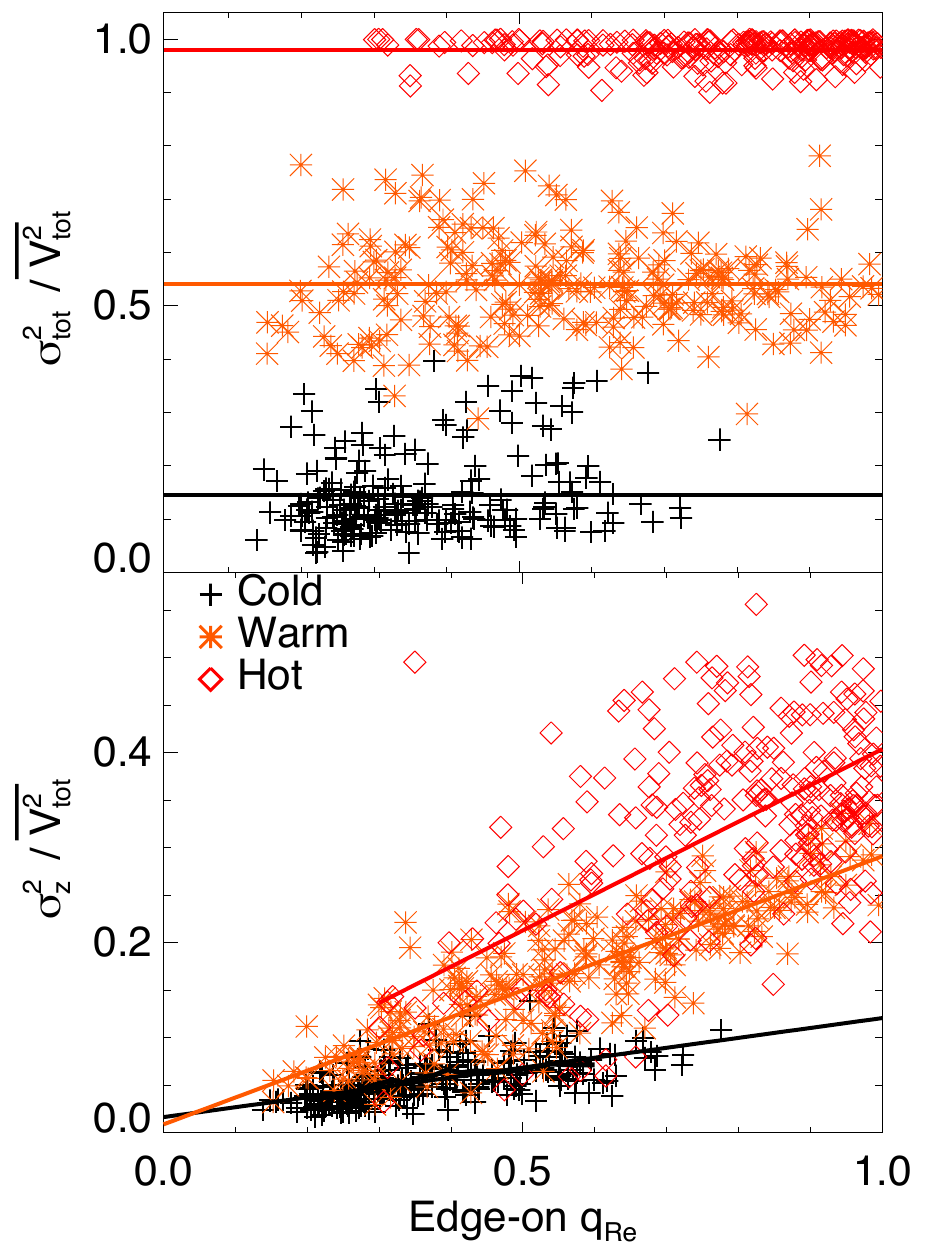}
\caption{ Velocity dispersion over total second velocity moment as function of intrinsic flattening.
{\bf Top:} $\sigma_{\mathrm{tot}}^2 / \overline{V_{\mathrm{tot}}^2}$ vs. intrinsic flattening $q_{\mathrm{Re}}$ for cold (black pluses), warm (orange asterisks) and hot (red diamond) components. The black, orange and red solid lines represent values of $0.145, 0.54, 0.98$, respectively. 
{\bf Bottom: } $\sigma_z^2 / \overline{V_{\mathrm{tot}}^2}$ vs. $q_{\mathrm{Re}}$. The black ($y = 0.017 + 0.103x$), orange ($y = 0.009 + 0.28x$) and red ($y = 0.022 + 0.38x$) solid lines are best linear fits to the cold, warm and hot components, respectively. }
\label{fig:sigmazt_qi90}
\end{figure}

Circularity is a good indicator of orbits' ordered-to-random motion, but it does not tell in which direction the random motion is dominated. To this end, we calculate the first order, second order velocity moments and intrinsic velocity dispersiosn for each component:
\begin{eqnarray}
\overline{V_k} = \frac{\int{V_{k}(x,y,z)f(x,y,z)}}{\int{f(x,y,z)}}
\\
\overline{V_k^2} = \frac{\int{V_k(x,y,z)^2f(x,y,z)}}{\int{f(x,y,z)}},
\\
\sigma_k^2 = \overline{V_k^2} - \overline{V_k}^2,
\end{eqnarray}
where $k$ represents the direction of ($R, \phi, z$) in cylindrical coordinates, $f(x,y,z)$ is the flux, and $V_k(x,y,z)$ are velocities at position (x,y,z). $\overline{V_k}$ represent the ordered rotation, while $\sigma_k$ represents the random motions along each direction. 
The total second moment is:
\begin{equation} 
\overline{V_{\mathrm{tot}}^2} = \overline{V_R^2} + \overline{V_z^2} + \overline{V_\phi^2},
\end{equation}
while the total random motion is:
\begin{equation} 
\sigma_{\mathrm{tot}}^2 = \sigma_R^2 + \sigma_z^2 + \sigma_{\phi}^2. 
\end{equation}

In Figure~\ref{fig:sigmazt_qi90}, we show $\sigma_{\mathrm{tot}}^2/\overline{V_{\mathrm{tot}}^2}$ vs. the intrinsic flattening $q_{\mathrm{Re}}$ in the top and $\sigma_{z}^2/\overline{V_{\mathrm{tot}}^2}$ vs. $q_{\rm Re}$ in the bottom. The black pluses, orange asterisks and red diamonds represent the cold, warm and hot components, respectively. 

Each component has an almost constant $\sigma_{\mathrm{tot}}^2/\overline{V_{\mathrm{tot}}^2}$ of $0.145, 0.54, 0.98$ for cold, warm and hot components, respectively. This is consistent with their definition of uniform circularity for each component. 

However, each component has a variation of $\sigma_{z}^2/\overline{V_{\mathrm{tot}}^2}$, that is strongly correlated with intrinsic flattening. Best linear fits of cold, warm, hot components are $\sigma_{z}^2/\overline{V_{\mathrm{tot}}^2} = 0.017 + 0.103q_{\mathrm{Re}}$, $\sigma_{z}^2/\overline{V_{\mathrm{tot}}^2} = 0.009 + 0.28q_{\mathrm{Re}}$, $\sigma_{z}^2/V_{\mathrm{tot}}^2 = 0.022 + 0.38q_{\mathrm{Re}}$, respectively. 

In the previous sections, we show that each component (cold/warm/hot) becomes thicker (larger $q_{\mathrm{Re}}$) in higher mass galaxies, which reflects the change of $\sigma_z^2/\overline{V_{\mathrm{tot}}^2}$.
This reveals a second order variation of orbit types; for each orbital component with constant circularity ($\sigma_{\mathrm{tot}}^2/\overline{V_{\mathrm{tot}}^2}$), the random motion ratio of $\sigma_z$ to $\sigma_{\mathrm{tot}}$ increases from low mass to high mass galaxies.

\subsection{Comparison with simulations}
\label{SS:NIHAO}
\begin{figure}
\centering\includegraphics[width=7.8cm]{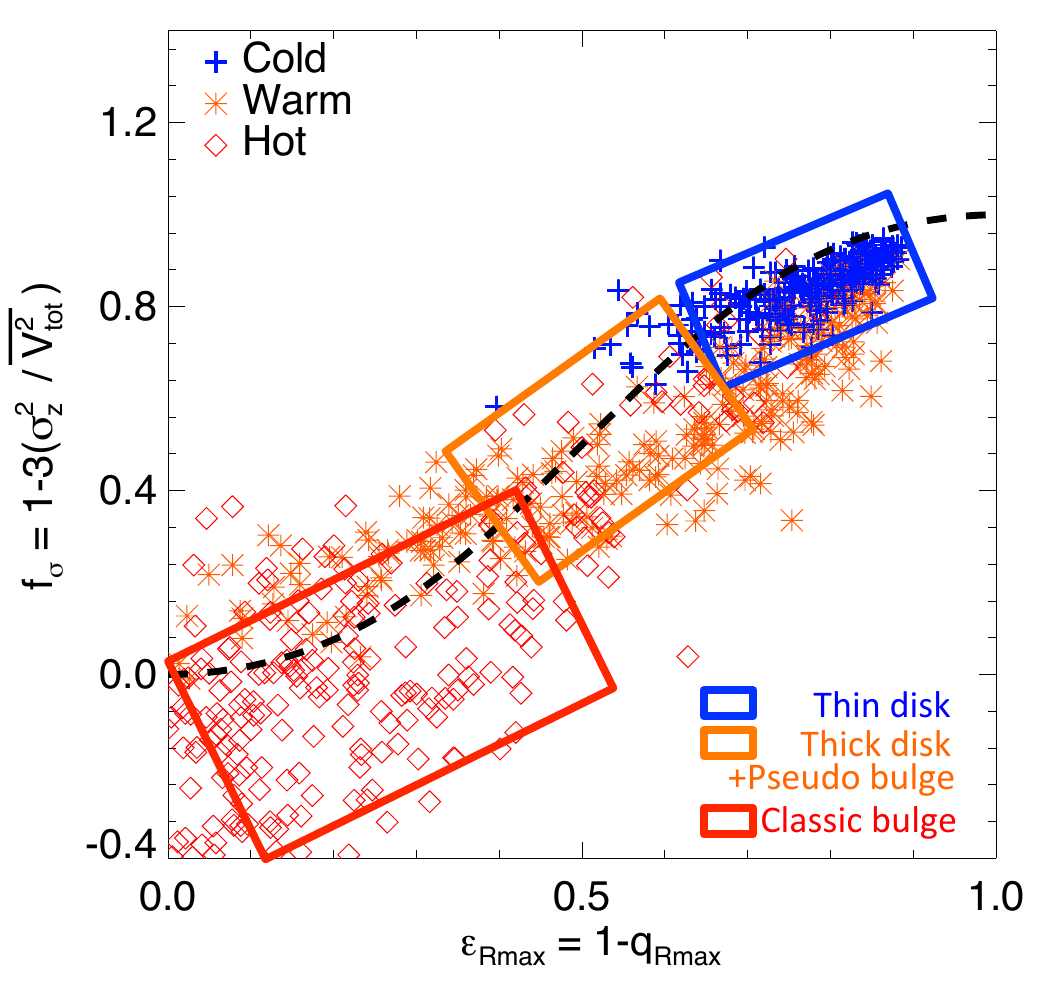}
\caption{The velocity dispersion fractions, $f_{\sigma}$  vs. $\epsilon_{\rm Rmax}$ of cold (blue pluses), warm (orange asterisks), and hot (red diamonds) components. The dashed black curve represents the function $f_{\sigma, \beta} = (1+(\epsilon / (1-\epsilon))^{-\beta})^{-1}$ with $\beta = 1.8$. The blue, orange and yellow squares indicate the regions occupied by thin disk, thick disk + pseudo bulge, and classic bulge identified from NIHAO simulations \citep{Obreja2018b}.}
\label{fig:sigmazt_2}
\end{figure}

\cite{Obreja2018b} kinematically identified multiple structures, for instance thin disk, thick disk, pseudo bulge, classic bulge, from 25 simulated galaxies from the NIHAO project \citep{Wang2015}.
The kinematics and shapes of these structures generally follow an analytic function: $f_{\sigma, \beta} = (1+(\epsilon / (1-\epsilon))^{-\beta})^{-1}$ with $\beta = 1.8$, where $\epsilon$ is the ellipticity and $f_{\sigma} = 1 - 3(\sigma_z^2/\overline{V_{\mathrm{tot}}^2})$ of each structure. The strongly rotating components follow the curve tightly, and components with weaker rotations have larger scatter. The thin disks in particular are clustered in a small region with $\epsilon\in [0.7,0.9]$ and $f_{\sigma} \in [0.8, 1.0]$. The thick disk and pseudo bulge lie roughly in the region of $\epsilon\in [0.4,0.7]$ and $f_{\sigma} \in [0.3, 0.8]$. 
The classic bulges have $\epsilon\in [0.0,0.5]$ and generally lie below the curve. The regions occupied by thin disk, thick disk + pseudo bulge, and classic bulge are indicated by the blue, orange and red squares in Figure~\ref{fig:sigmazt_2}.

We calculate similar properties for the orbital components. $f_{\sigma}$ versus $\epsilon_{\rm Rmax}$ of cold, warm, hot components are represented by blue pluses, orange asterisks, and red squares in Figure~\ref{fig:sigmazt_2}, where $\epsilon_{\rm Rmax} = 1 - q_{\rm Rmax}$ is our best comparison with $\epsilon$. The black dashed curve represents the analytic function $f_{\sigma, \beta}$. 

Our cold, warm, and hot components generally follow the relation defined by $f_{\sigma, \beta}$. The cold components follow the relation tightly, occupying similar regions as the thin disk of NIHAO galaxies. The warm components have larger scatters, span the region of thick disk + pseudo bulge, and extend to other regions along the $f_{\sigma, \beta}$ curve. The hot components mainly occupy the region with small $\epsilon$ and lie below the curve, similar to the classic bulges. 

In previous sections, we shown that the orbital components change their morphology, from disk-like in low-mass late-type galaxies to bulge-like in high-mass early-type galaxies. The spread of warm and hot components along $f_{\sigma, \beta}$ curve again reveals their variety of kinematics as well in our sample. 

\section{discussion}
\label{S:discussion}
Two distinct mechanics are possible to explain the low angular momentum of galaxies in present-day universe: (i) heating from the cold disk, e.g., secular evolution or galaxy mergers and (ii) rapid star formation in early violent universe \citep{Lagos2017}. Both channels lead to loss of angular momentum, resulting in more compact and rounder galaxies (e.g.,\citealt{Lagos2018}, \citealt{Obreja2018b}), in agreement with the known correlation between a galaxy's global morphology and its kinematics.
However, global galaxy properties alone are not enough to distinguish the effect of these two channels in different galaxies.  


In \citet{Zhu2018b}, the orbit's circularity distributions are further derived and divided into four components. We showed that high circularity cold orbits decreases fast, while random motion dominated hot orbits increases, and the warm components keep almost constant in galaxies from $10^{10}$ to $10^{11}\, M_{\odot}$. Of course these are consistent with the decrease of total angular momentum. But it includes more information than that. Stars on cold, warm and hot orbits may have totally different formation paths: we understand from the simulations that stars formed in-situ have increasing angular momentum in the more recent universe; with thicker disk forming earlier and thin disk forming later (e.g., \citealt{Brook2004}; \citealt{Bird2013}; \citealt{Stinson2013}).
The cold components are mostly young stars formed in-situ, warm component could trace old stars formed in-situ, or stars being heated from cold disks via secular evolution, and a small fraction of warm component could be accreted \citep{Gomez2017}.
The stars on hot orbits should be mostly accreted (\citealt{Gomez2017}; \citealt{Tissera2017}) via minor or major mergers.
Further comparison with simulations will help us understanding the physical processes that may lead to the present-day orbit distribution. 
We notice that in galaxies with $M>10^{11}\, M_{\odot}$, the warm components also decrease sharply with hot components become dominating, which is a clear indication of a different formation channel. Major dry mergers could be responsible for this result \citep{Hoffman2010}.


Here we further show that the correlation between morphology and angular momentum results from combination of two effects: (i) components with lower angular momentum are generally rounder and more radially concentrated than components with higher angular momentum and (ii) each component (even with similar circularity) becomes significantly rounder (larger flattening $q_{\mathrm{Re}}$ and $q_{\mathrm{Rmax}}$) and more radially concentrated (larger Sersic index \textit{n} and concentration $C$) in higher mass galaxies. 
The first is just what we expected. While the latter indicates that morphology could actually be a secondary independent fossil record of the galaxy formation history. Just as shown in Section~\ref{SS:orbit}, flattening brings in new information of the random motion direction.  
Concentration could reflect the density of environment when the galaxy was formed.  
We need simulations to study in detail the similar properties of galaxies formed from different types of mergers and rapid star formation in early universe, before we can tell the formation path of the real galaxies based on these information.

Also with these complexities between the morphology and kinematics, the photometric-identified bulges are usually combination of different orbital components as we shown in Section~\ref{SS:photmetric}, thus should include stars from multiple origins \citep{Bell2017}. We believe a combination of stellar orbit distribution and its morphological properties will help to further disentangle these complicated physical processes in galaxy formation history.

\section{Summary and conclusion}
\label{S:summary}
We started from stellar orbit distributions obtained by orbit-based Schwarzschild models for a sample of 250 galaxies across the Hubble sequence. Based on the circularity, we decompose the galaxies into four orbital components: cold with strong rotation ($\lambda_z = 0.92\pm0.02$), warm with weak rotation ($\lambda_z = 0.54\pm0.05$), hot with negligible rotation ($\lambda_z = 0.02\pm0.05$) and a counter-rotating components ($\lambda_z = -0.53\pm0.12$).
We then rebuild the surface brightness maps of each orbital component and quantify their morphology with Sersic index \textit{n}, concentration $C = \log{(\Sigma_{0.1R_e}/\Sigma_{R_e})}$ and intrinsic flattening $q_{\mathrm{Re}}$ and $q_{\mathrm{Rmax}}$ with $R_{\mathrm{max}}$ the CALIFA data coverage. Comparing these results with 
photometric-identified bulge and disk from observations and kinematic-identified structures from simulations, we find: 

\begin{itemize}
\item In general, kinematic hotter components are more spatially concentrated (with larger Sersic index \textit{n} and concentration $C$) and rounder (with larger intrinsic flattening $q_{\rm Re}$ and $q_{\rm Rmax}$) than colder components. From the surface brightness maps projected edge-on, we get $n = 1.0\pm0.8, 2.0\pm1.5, 2.7\pm1.6, 2.0\pm2.0$, and $C = 0.6\pm0.3, 1.1\pm0.5, 1.5\pm0.5, 1.6\pm1.1$, for cold, warm, hot and CR components respectively. The intrinsic flattening are measured within half-light-radius $R_e$ and the CALIFA data coverage $R_{\mathrm{max}}$ ($R_{\mathrm{max}} \sim 2 R_e $). The average values are $q_{\mathrm{Re}} = 0.37\pm0.14, 0.57\pm0.21, 0.77\pm0.18, 0.57\pm0.23$ and $q_{\mathrm{Rmax}} = 0.23\pm0.10, 0.44\pm0.22, 0.67\pm0.22, 0.48\pm0.25$ from cold, warm, hot, and CR components, respectively.

\item All components (cold/warm/hot) become significantly more concentrated and rounder in higher mass galaxies. We quantify their variations as function of total stellar mass $M_*$. From the radial profiles, we get $n = (-14.1\pm1.8) + (1.5\pm0.2)\times \log10(M_*)$, $C = (-5.4\pm0.7) + (0.61\pm0.07)\times \log10(M_*)$ for the warm component, and $n = (-12.2\pm3.5) + (1.4\pm0.3)\times \log10(M_*)$, $C = (-1.7\pm0.9) + (0.32\pm0.08)\times \log10(M_*)$ for the hot component. From the intrinsic flattening, we get $q_{\rm Re} = (-1.85\pm0.39) + (0.23\pm0.04)\times \log10(M_*)$ for the warm component and $q_{\rm Re} = (-0.02\pm0.23) + (0.07\pm0.02)\times \log10(M_*)$ for hot component. 
Cold components are almost all disk-like but become thicker in more massive galaxies. Warm components change from disk-like to bulge-like from low-mass late-type galaxies to high-mass early-type galaxies. Hot components are mostly bulge-like, but could become disk-like in low-mass Sc+Sd galaxies.

\item The cold orbits luminosity fraction is well correlated with photometrically decomposed disk fraction with $f_{\mathrm{cold}} = (0.14\pm0.05) + (0.23\pm0.09)f_{\mathrm{\mathrm{disk}}}$. Hot orbits are correlated with bulges with $f_{\mathrm{hot}} = (0.19\pm0.04) + (0.31\pm0.05)f_{\mathrm{\mathrm{bulge}}}$. Warm orbits mainly contribute to disks in low-mass late-type galaxies, and to bulges in high-mass early-type galaxies. 

\item The significant overlap of warm and hot components in Sersic \textit{n} suggests that Sersic \textit{n} is not a good separation for rotating and non-rotating bulges, although at any fixed stellar masses, bulges with higher Sersic \textit{n} should still generally have weaker rotations. 

\item The above scenario is consistent with their intrinsic flattening distributions comparing to those of disks and bulges from other images studies. The intrinsic flattening distribution of spiral galaxies and elliptical galaxies we obtained are consistent with those from statistic study of galaxy images \citep{Rodrguez2013}. While for the spiral galaxies having bulge and disk, the cold components are generally thinner than the disks, warm components varies from slightly thicker than disks to be really round, hot components span the similar range as the ellipticals and the intrinsic flattening of bulges from \citet{Costantin2018}. 

\item The increasing of intrinsic flattening with stellar mass reflects the increase of random motion at $z$ direction; $\sigma_z^2/\overline{V_{\mathrm{tot}}^2}$ of each component increases with galaxy' stellar mass. It is a secondary relation with $M_*$ independent of circularity distribution itself, which may further constrain the galaxies' formation histories. Our cold, warm, and hot components generally follow the same morphology ($\epsilon = 1-q_{\rm Rmax}$) versus kinematics ($\sigma_z^2/\overline{V_{\mathrm{tot}}^2}$) relation as the thin disk, thick disk + pseudo bulge, and classic bulge identified from NIHAO galaxies. 

\item The surface brightness profiles of the four components we present could potentially be used as templates in photometric decomposition for separating similar components from images.
\end{itemize}


\bibliography{kin} 
\section*{Acknowledgment}
GvdV acknowledges funding from the European Research Council (ERC) under the European Union's Horizon 2020 research and innovation programme under grant agreement No 724857 (Consolidator Grant ArcheoDyn). JMA acknowledge  support  from  the Spanish  Ministerio  de Economia y Competitividad (MINECO) by the grant AYA2013-43188-P, AO has been funded by the Deutsche Forschungsgemeinschaft (DFG, German Research Foundation) -- MO 2979/1-1.

\appendix
\section{Model uncertainties}
\label{S:error}
\subsection{Model recovery of the morphology}
We applied the Schwarzschild model to 131 mock data created from 14 cosmological simulated galaxies, 5 from NIHAO \citep{Wang2015}, 5 from Auriga \citep{Grand2017} and 4 from Illustris \citep{Illustris2014}. There are 2 low-mass spiral galaxies with $m_* < 10^{10}\,M_{\odot}$, 8 massive spiral galaxies with $m_* > 10^{10}\,M_{\odot}$ and 4 early-type galaxies.
From each simulation, 7 or 10 sets of mock data are created with different viewing angles of the projection. The inclination angle range of $30^o-90^o$ are uniformly covered by the 131 mock data sets. The mock data have similar quality to the CALIFA data, in the sense of similar spatial resolution,  with only mean velocity and velocity maps (not $h_3$ or $h_4$) and similar errors of the kinematic maps.

We have shown the model recovery of luminosity fractions of cold, warm, hot, and CR components and evaluated the systematic bias and errors of these luminosity fractions according to these tests (see details in \citet{Zhu2018b}).
Here we use the same models to quantify the recovery of morphology of different orbital components. 
For each simulation, we first decompose it into cold, warm, hot and CR components based on its true orbit distribution. We measure the true values of Sersic index $n_{\rm true}$, concentration $C_{\rm true}$, intrinsic flattening $q_{\rm Re, true}$ and $q_{\rm Rmax, true}$, the same way as we described in Section~\ref{S:morphology}.

Each of the 131 mock data sets is taken as an observed galaxy. With the best-fitting Schwarzschild model, we decompose the galaxy into cold, warm, hot and CR components due to the orbit distribution we obtained. The morphology of each component, Sersic index $n_{\rm model}$, concentration $C_{\rm model}$, intrinsic flattening $q_{\rm Re, model}$ and $q_{\rm Rmax, model}$ are measured.

\begin{figure}
\centering\includegraphics[width=8.0cm]{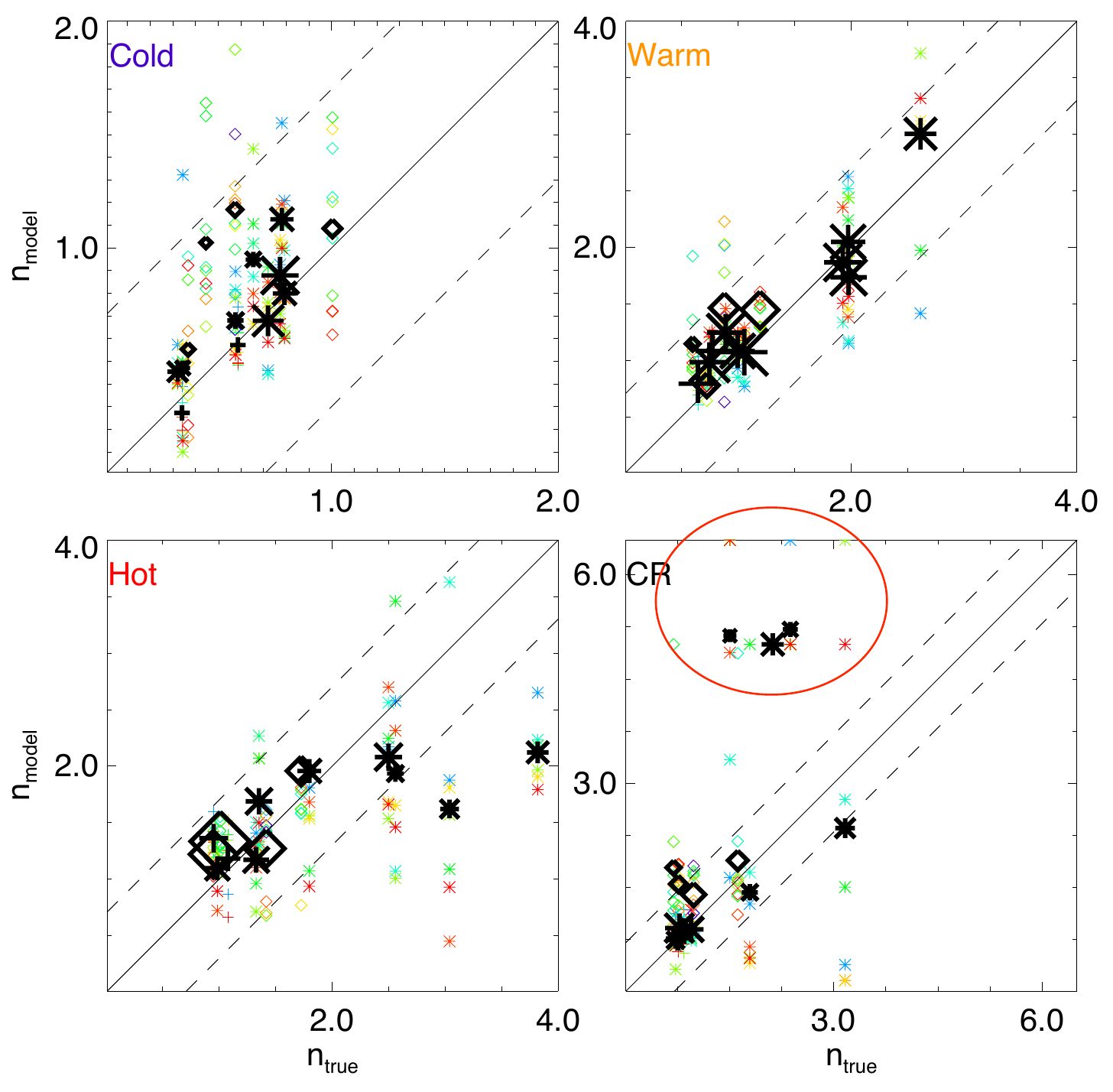}
\caption{One-to-one comparison of Sersic $n_{\rm model}$ vs. $n_{\rm true}$ of the 131 mock data sets. The four panels are for cold, warm, hot and CR components, respectively. In each panel, one colored dot represent one mock data sets, with colors indicating the inclination angles, from face-on to edge-on with colors from blue to red. Each black symbol represents the average $n_{\rm model}$ obtained by the 7/10 mock data from the same simulation. The sizes of the black symbols are scaled with luminosity fractions of the corresponding orbital components, pluses represent low-mass spiral galaxy, asterisks represent massive spiral galaxies and the diamonds represent early-type galaxies. The solid line is one-to-one, the dash lines represent shifts of $\pm0.6$. The points for CR in the red circle are obviously highly biasd points, which we can select out (and exclude) in real galaxies as we did in Section~\ref{SS:SB}.  The points in the red circles are not included in evaluating the systematic uncertainties.}
\label{fig:n_one-to-one}
\end{figure}

\begin{figure}
\centering\includegraphics[width=8.0cm]{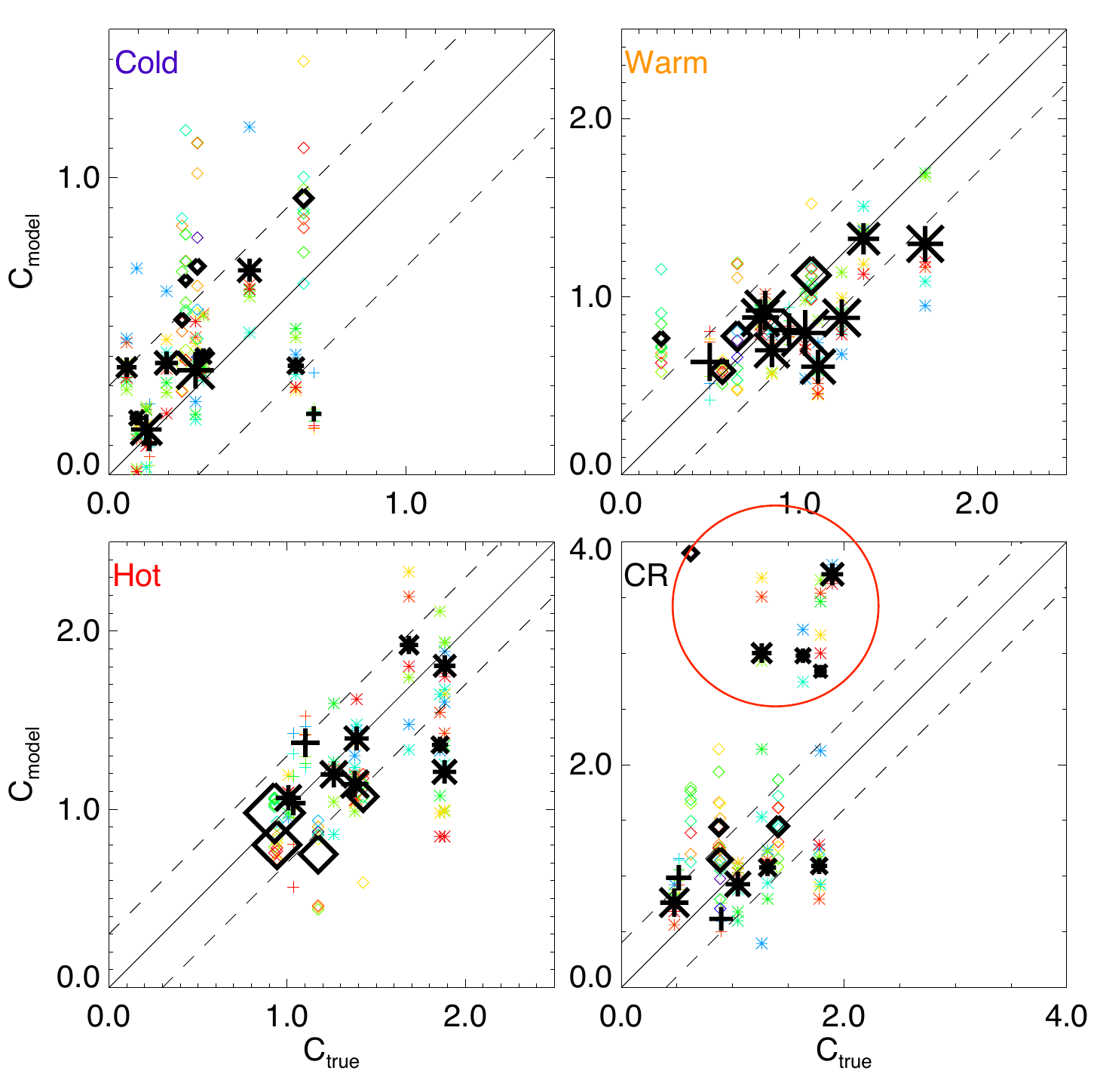}
\caption{Similar to Figure~\ref{fig:n_one-to-one}, but for concentration $C$. The dashed line is one-to-one, the dashed lines represent shifts of $\pm0.3$. The points in the red circles are not included in evaluating the systematic uncertainties. }
\label{fig:C_one-to-one}
\end{figure}

The one-to-one comparison of Sersic $n_{\rm model}$ vs. $n_{\rm true}$ of the four components are shown in Figure~\ref{fig:n_one-to-one}. $n_{\rm model}$ matches $n_{\rm true}$ generally well, except a few points in hot and CR components. $n$ of the hot component are under-estimated in two spiral galaxies which have small hot orbit fractions. For the CR components, we get some biased points which also have small CR orbit fractions.  The biased points of CR components towarding very high $n_{\rm model}$ are circled out.
Similar comparison of concentration $C$ is shown in Figure~\ref{fig:C_one-to-one}.  In the CR components, there are some similar biased points toward very high $C_{\rm model}$ which we circled out.
We consider the biased points circled out are features we can pick out for real galaxies. As we did in Section~\ref{SS:SB}, the points with $n>6.5$ or $C>3$ are excluded from the analysis.

\begin{figure}
\centering\includegraphics[width=8.0cm]{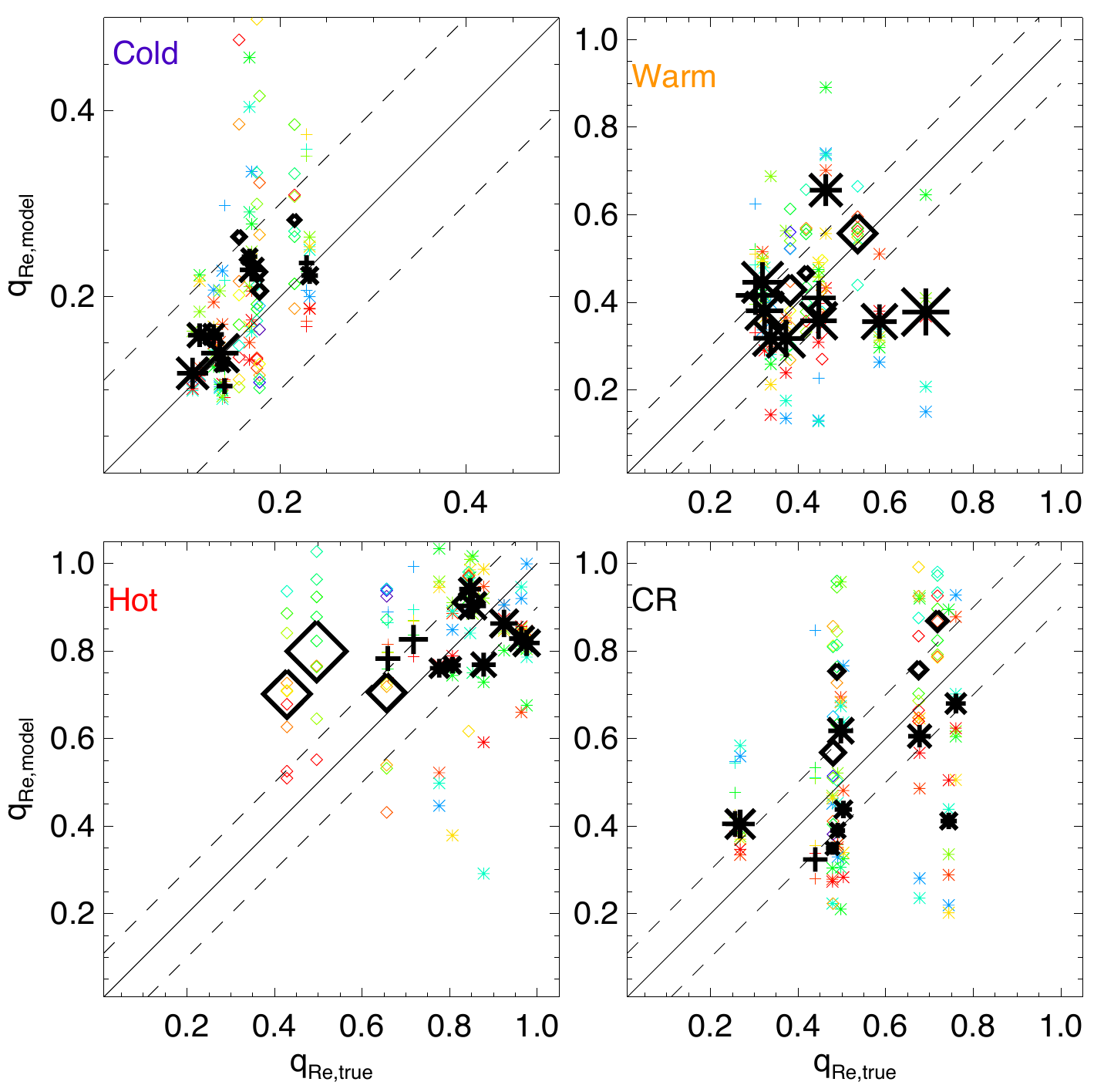}
\caption{Similar to Figure~\ref{fig:n_one-to-one}, but for intrinsic flattening $q_{\rm Re}$. The dashed line is one-to-one, the dashed lines represent shifts of $\pm0.1$.}
\label{fig:q_one-to-one}
\end{figure}

\begin{figure}
\centering\includegraphics[width=8.0cm]{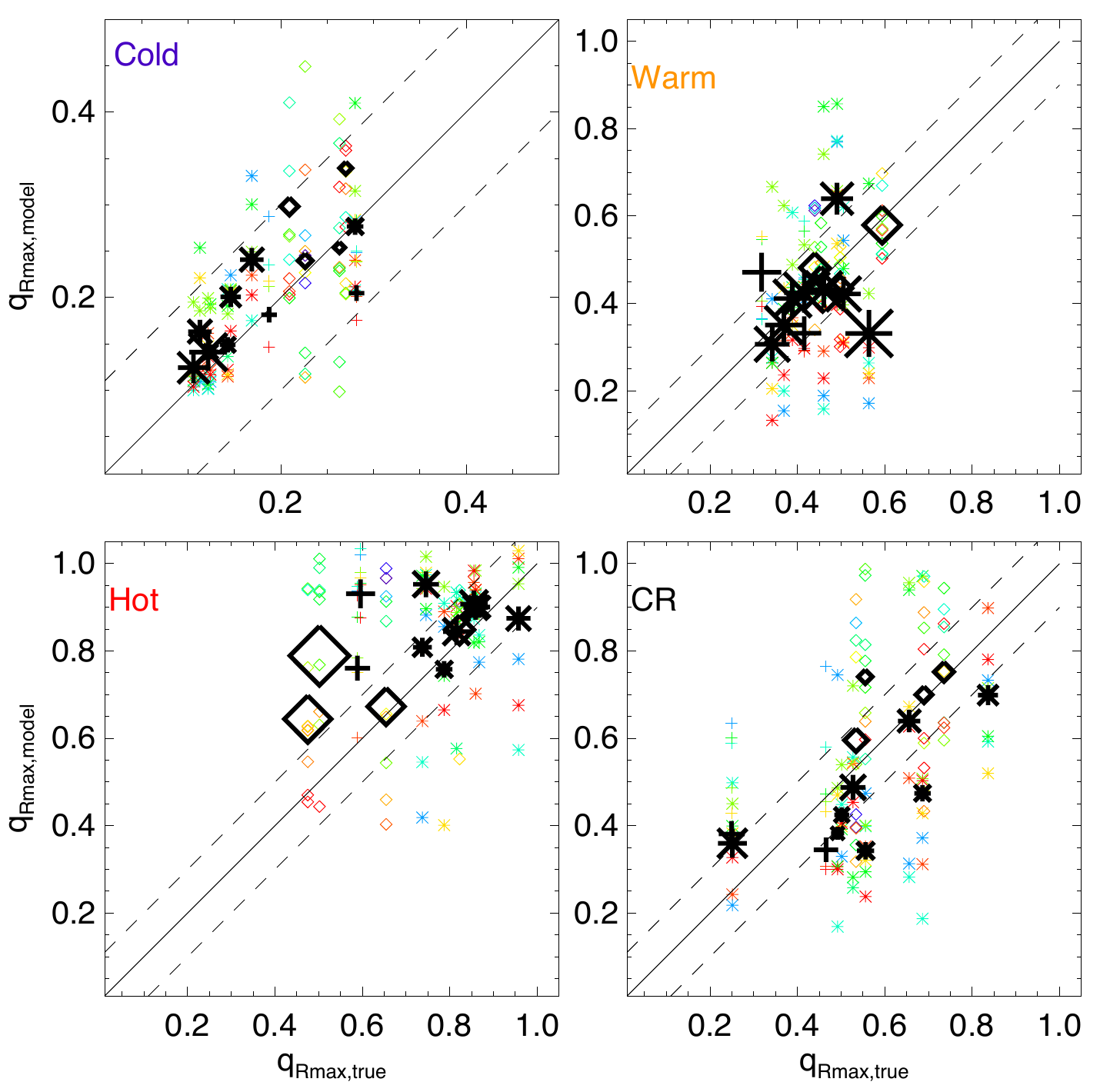}
\caption{Similar to Figure~\ref{fig:n_one-to-one}, but for intrinsic flattening $q_{\rm Rmax}$. The dashed line is one-to-one, the dashed lines represent shifts of $\pm0.1$.}
\label{fig:qRmax_one-to-one}
\end{figure}

The similar model recovery of intrinsic flattening $q_{\rm Re}$ and $q_{\rm Rmax}$ are shown in Figure~\ref{fig:q_one-to-one} and Figure~\ref{fig:qRmax_one-to-one}. The models generally match the true values of the simulations, with some scatters. Note that for the intrinsic flattening, there is not significantly larger bias for the components with small fractions. 

As we have included different viewing angles in the range of $30^o-90^o$ in the mock data, generally representative of the CALIFA sample, 
and there is no significant preference of inclination angles revealed from Figure~\ref{fig:n_one-to-one} to Figure~\ref{fig:qRmax_one-to-one},  we do not consider the difference with viewing anlges when evaluating the systematic uncertainties. 

We evaluate systematic bias and systematic errors by dividing the 14 galaxies into 3 groups, low-mass spiral galaxies, high-mass spiral galaxies and early-type galaxies. To summarize the model uncertainties for different types of galaxies, we define deviation $\mathcal{D} = p_{\rm model} - p_{\rm true}$ from each single mock data sets, where $p$ represents any of $C$, $n$, $q_{\rm Re}$ and $q_{\rm Rmax}$ for any of the four orbital components. 
The average and standard deviation of relative deviation are calculated with
$\overline{d} = \overline{p_{\rm model} - p_{\rm true}}/\overline{p_{\rm model}}$ and $\sigma(d) = \sigma(p_{\rm model} - p_{\rm true})/\overline{p_{\rm model}}$ for each type of galaxies. For Sersic index $n$ and $C$ of the CR component, we calculate $\overline{d}$ and $\sigma(d)$ by excluding these points circled out. As similar points are also excluded in the analysis in Section~\ref{SS:SB}, we consider $\overline{d}$ and $\sigma(d)$ calculated here represent the uncertainties of the points kept in the analysis.

In table~\ref{tab:error}, we list $\overline{d}$ and $\sigma(d)$, for $C$, $n$, $q_{\rm Re}$ and $q_{\rm Rmax}$ of the four orbital components. Then the systematic bias and error for parameter $p$ are just $\overline{d_p} p$ and $\sigma(d_p) p$. We assign errors to each of CALIFA galaxies according to its galaxy type, and these errors are used in Section~\ref{S:morphology}.

\subsection{Model recovery of the morphology-kinematic relation}
\begin{figure}
\centering\includegraphics[width=8.0cm]{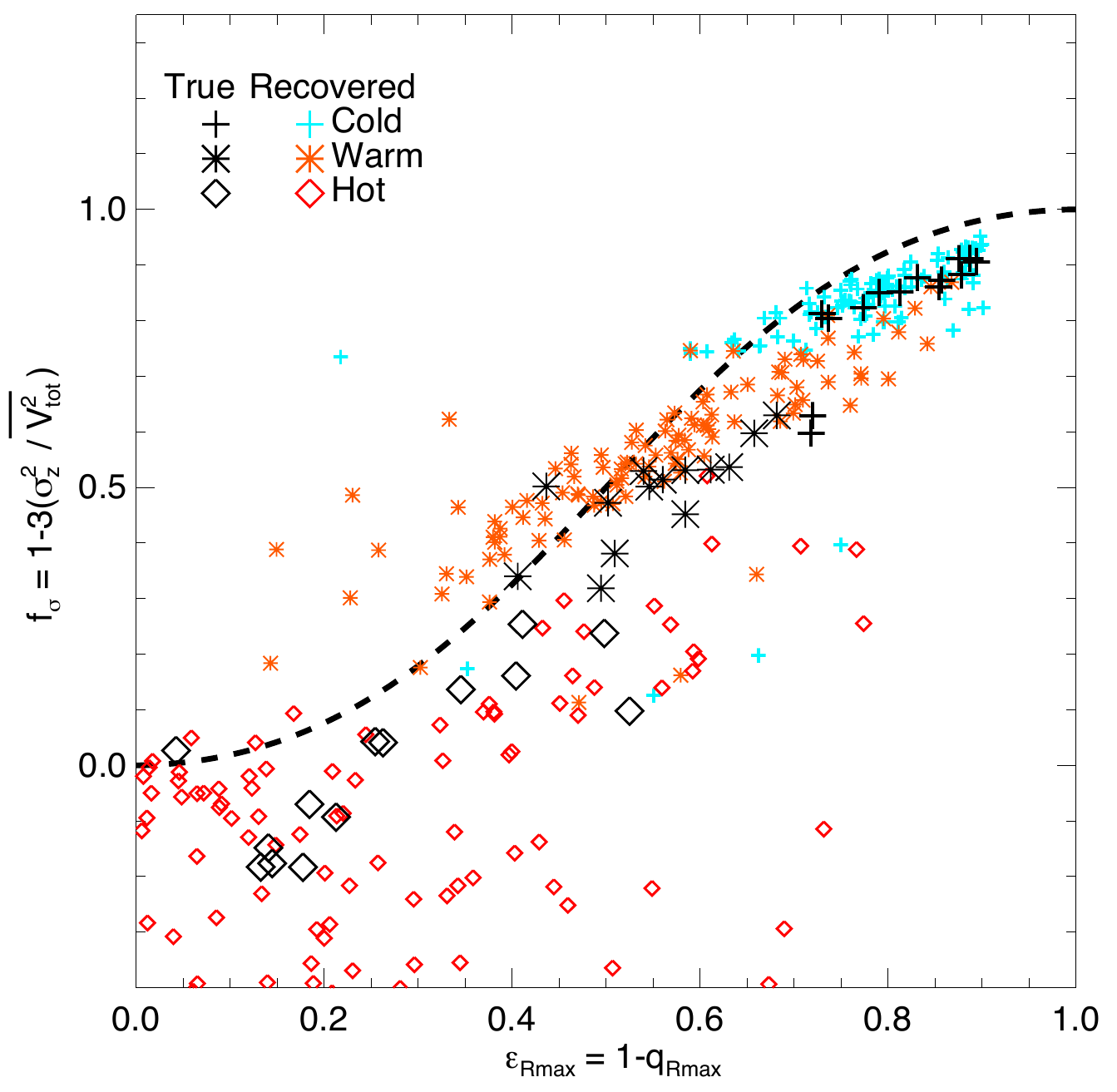}
\caption{Model recovery of the morphology-kinematic relation, $f_{\sigma}$ versus. $\epsilon$. With the 14 simulated galaxies, the black pluses, asterisks, diamonds represent the true values of cold, warm, and hot components. The true values are directly calculated from the particles in the simulations, each point represents one simulated galaxy. The blue pluses, orange asterisks, and red diamonds are what recovered from our model for cold, warm, and hot components, respectively, each points represents that obtained from one mock-data set. The dashed black curve represents the function $f_{\sigma, \beta} = (1+(\epsilon / (1-\epsilon))^{-\beta})^{-1}$ with $\beta = 1.8$. }
\label{fig:Sz_qRmax_recover}
\end{figure}

We also calculate $\sigma_{z}^2/\overline{V_{\mathrm{tot}}^2}$ of the cold, warm, and hot component from the simulations directly. 
$f_{\sigma} = 1-3\sigma_{z}^2/\overline{V_{\mathrm{tot}}^2}$ versus $\epsilon$ of the 14 simulations are shown as black points in Figure~\ref{fig:Sz_qRmax_recover}. The colored points are those recovered from our model with the 131 mock data sets created from those simulations. The dashed black curve represents the function $f_{\sigma, \beta} = (1+(\epsilon / (1-\epsilon))^{-\beta})^{-1}$ with $\beta = 1.8$. Our model recovered (colored) points follow the analytic curve similarly to the (black) points of the true values, although our model recovered points have slightly larger scatters. The cold, warm, and hot components occupy different regions on the $f_{\sigma}-\epsilon$ diagram, and our model recovered values follow those of the true values as well.

\begin{table*}
\caption{Relative systematic bias and errors of the parameters $p$ representing any of $C$, $n$, $q_{\rm Re}$ and $q_{\rm Rmax}$. $\overline{d} = \overline{p_{\rm model} - p_{\rm true}}/\overline{p_{\rm model}}$ and $\sigma(d) = \sigma(p_{\rm model} - p_{\rm true})/\overline{p_{\rm model}}$ are obtained for cold, warm, hot and CR components for each of the three groups of galaxies.}
\label{tab:error}\footnotesize
\begin{tabular}{*{13}{l}}
\hline
Param & \multicolumn{4}{|c|}{Low-mass late-type} &  \multicolumn{4}{c|}{High-mass late-type} & \multicolumn{4}{c|}{Early-type} \\
\hline
-     & cold & warm & hot & CR					& cold & warm & hot & CR					& cold & warm & hot & CR				  \\
\hline
$\overline{d_n}$ & -0.10  &  0.24  &  0.20  &  0.02  &  0.18  &  0.06  & -0.27  &  0.38  &  0.37  &  0.29  &  0.12  &  0.38 \\
$\sigma(d_n)$ &    0.20  &  0.13  &  0.17  &  0.22  &  0.28  &  0.23  &  0.37  &  0.33  &  0.35  &  0.22  &  0.32  &  0.44 \\
\hline
$\overline{d_C}$ &   -2.04  &  0.01  &  0.11  &  0.12  &  0.24  & -0.20  & -0.11  &  0.28  &  0.39  &  0.08  & -0.25  &  0.15 \\
$\sigma(d_C)$ & 0.51  &  0.15  &  0.17  &  0.13  &  0.37  &  0.15  &  0.17  &  0.32  &  0.33  &  0.20  &  0.16  &  0.22 \\
\hline
$\overline{d_{qRe}}$& 0.15  &  0.15  &  0.19  &  0.33  &  0.16  & -0.11  & -0.06  & -0.13  &  0.24  &  0.04  &  0.25  &  0.19 \\
$\sigma(d_{qRe})$ &  0.49  &  0.22  &  0.07  &  0.35  &  0.30  &  0.31  &  0.16  &  0.35  &  0.47  &  0.19  &  0.19  &  0.20 \\
\hline
$\overline{d_{qRmax}}$ & -0.07  &  0.18  &  0.36  &  0.35  &  0.17  & -0.05  &  0.04  & -0.19  &  0.14  &  0.00  &  0.20  &  0.10 \\
$\sigma(d_{qRmax})$ &    0.21  &  0.25  &  0.11  &  0.38  &  0.24  &  0.37  &  0.15  &  0.33  &  0.45  &  0.16  &  0.24  &  0.26 \\
  \hline
 \end{tabular}
\end{table*}

\onecolumn
\begin{tiny}
\LTcapwidth=\textwidth
\begin{landscape}

\begin{longtable}{*{30}{l}}
\caption{The luminosity fractions and morphological parameters of cold, warm, hot and CR components of 260 CALIFA galaxies. The first nine columns are galaxy name, Hubble type, stellar mass $\log10(M_*/M_{\odot})$ from \citet{Walcher2014}, SDSS magnitudes $M_g$, $M_r$, $M_i$, volume of the galaxies (volume calculated with the 960 galaxies as mother sample of CALIFA from \citet{Walcher2014}), half-light-radii $R_e$ in kpc, the CALIFA data coverage $R_{\rm max}$ in kpc, following 4 columns for each of the cold, warm, hot, and CR components, they are luminosity fractions within $R_e$ $f_{\rm R_e}$, within 5kpc $f_{\rm 5}$, intrinsic flattening $q_{\rm Re}$, concentration $C$ and Sersic index $n$. 250 of the 260 galaxies have $R_{\rm max} > R_e$ as we used for analysis in the paper, 246 of 260 galaxies have $R_{\rm max} > 5$ kpc. The galaxy names, MCG-01-54-016, MCG-02-02-040, MCG-02-02-030, MCG-02-03-015, MCG-02-51-004 have been shorten by excluding 'MCG-', UGC11680NED01 has been shorten to be NED01. For those components not having a reason Sersic fit or not bright enough for measuring $q_{\rm Re}$ or $C$, the corresponding parameters are set to be zeros. The full Sersic fitting parameters for each galaxy are only included in the online material.}\\

\hline
\label{tab:orbitf}
&&&&& & & & & \multicolumn{5}{|c|}{Cold} & \multicolumn{5}{c|}{Warm} & \multicolumn{5}{c|}{Hot} & \multicolumn{5}{c}{CR} \\
\hline
Galaxy & Type & $M_*$ & $M_g$& $M_r$ & $M_i$ & volume & $R_e$ & $R_{\rm max}$ &  $f_{\rm R_e} $ & $f_{\rm 5}$ & $q_{\rm Re}$ & $C$ & $n$  &  $f_{\rm R_e} $ & $f_{\rm 5}$ & $q_{\rm Re}$ & $C$ & $n$   & $f_{\rm R_e} $ & $f_{\rm 5}$ & $q_{\rm Re}$ & $C$ & $n$    & $f_{\rm R_e} $ & $f_{\rm 5}$ & $q_{\rm Re}$ & $C$ & $n$ \\
\hline
UGC04722 &           Sdm &   8.68 & -18.05 & -18.18 & -18.37 &    7865 &   3.83 &   4.65 & 0.21 & 0.14 & 0.15 & 0.00 & 0.00 & 0.39 & 0.45 & 0.20 & 0.00 & 2.00 & 0.27 & 0.32 & 0.31 & 0.00 & 0.37 & 0.13 & 0.09 & 0.45 & 0.00 & 0.37 \\
\hline
\hline
\end{longtable}

\end{landscape}
\end{tiny}
\twocolumn

\end{document}